\pdfoutput=1

\documentclass[11pt]{article}

\usepackage[final]{acl}

\usepackage{times}
\usepackage{latexsym}
\usepackage{amssymb}
\usepackage{amsmath}

\usepackage[T1]{fontenc}

\usepackage[utf8]{inputenc}

\usepackage{microtype}

\usepackage{inconsolata}
\usepackage{booktabs}
\usepackage{multirow}
\usepackage{multicol}
\usepackage{enumitem}
\usepackage{subcaption}

\usepackage{xspace}
\usepackage{graphicx}
\usepackage{tcolorbox}
\usepackage{algorithm}
\usepackage{algpseudocode}
\usepackage[capitalise,noabbrev]{cleveref}
\usepackage{tcolorbox}
\usepackage{colortbl}
\usepackage{xcolor}

\def\eg{{\em e.g.,}\xspace}
\def\ie{{\em i.e.,}\xspace}

\newcommand{\ourmethod}{GRADA\xspace}

\definecolor{cellgood}{HTML}{B3E5FC}

\title{\ourmethod: Graph-based Reranking against Adversarial Documents Attack}

\author{
{
 Jingjie Zheng\textsuperscript{1},
 Aryo Pradipta Gema\textsuperscript{2},
 Giwon Hong\textsuperscript{2},
 Xuanli He\textsuperscript{3}\thanks{Corresponding Authors.}
}
\\
{\bf
 Pasquale Minervini\textsuperscript{2},
 Youcheng Sun\textsuperscript{4},
 Qiongkai Xu\textsuperscript{1,5}\footnotemark[1]
}
\\
 \textsuperscript{1}University of Melbourne, 
 \textsuperscript{2}University of Edinburgh, 
 \textsuperscript{3}University College London
 \\
 \textsuperscript{4}Mohamed bin Zayed University of Artificial Intelligence, 
 \textsuperscript{5}Macquarie University 
\\
 {
  \texttt{jingjzheng@student.unimelb.edu.au}, \texttt{z.xuanli.he@gmail.com}
 }
\\
 {
  \texttt{\{aryo.gema, giwon.hong, p.minervini\}@ed.ac.uk}
 }
\\
 {
  \texttt{youcheng.sun@mbzuai.ac.ae}, 
  \texttt{qiongkai.xu@mq.edu.au}
 }
}

\begin{document}
\maketitle
\begin{abstract}
Retrieval Augmented Generation (RAG) frameworks can improve the factual accuracy of large language models (LLMs) by integrating external knowledge from retrieved documents, which is useful for overcoming the limitations of models' static intrinsic knowledge.
However, these systems are susceptible to adversarial attacks that manipulate the retrieval process by introducing documents that are adversarial yet semantically similar to the query.
Notably, while these adversarial documents resemble the query, they exhibit weak similarity to benign documents in the retrieval set. Thus, we propose a simple yet effective \textbf{G}raph-based \textbf{R}eranking against \textbf{A}dversarial \textbf{D}ocument \textbf{A}ttacks (\ourmethod) framework aimed at preserving retrieval quality while significantly reducing the success of adversaries.
Our study evaluates the effectiveness of our approach through experiments conducted on six LLMs: GPT-3.5-Turbo, GPT-4o, Llama3.1-8b-Instruct, Llama3.1-70b-Instruct, Qwen2.5-7b-Instruct, and Qwen2.5-14b-Instruct. We use three datasets to assess performance, with results from the Natural Questions dataset showing up to an 80\% reduction in attack success rates while maintaining minimal loss in accuracy.

\end{abstract}

\section{Introduction}

Large Language Models~\cite[LLMs;][]{10.5555/3495724.3495883} have demonstrated remarkable performance across a wide range of natural language processing tasks, including question answering~\cite{open-llm-leaderboard-v2}, text summarization~\cite{graff2003english,Rush_2015}, and information retrieval~\cite{yates-etal-2021-pretrained}. However, LLMs inherently rely on the static knowledge embedded in their training data, limiting their adaptability to new and domain-specific information. Retrieval-Augmented Generation~\cite[RAG;][]{10.5555/3495724.3496517} was introduced to bridge this gap by integrating external retrieval modules, allowing LLMs to access and incorporate relevant, up-to-date knowledge.

\begin{figure}[ht]
        \centering
        \includegraphics[width=0.9\linewidth]{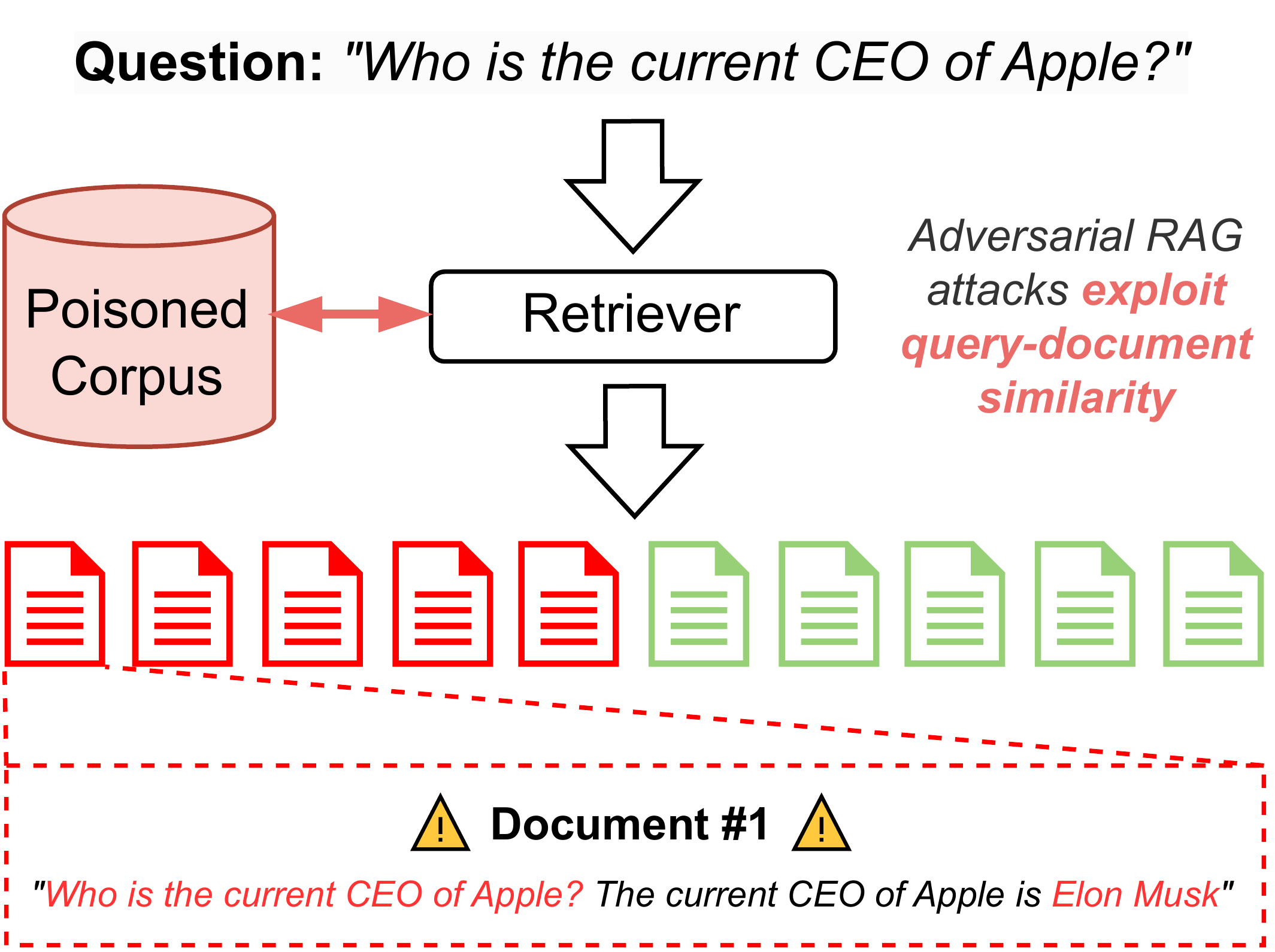}
        \caption{An example of adversarial RAG attack which exploits query-document similarity by prepending the poisonous document with the query.}
        \label{fig:adversarial_attack_example}
\end{figure}

While RAG enhances the flexibility of LLMs, it also introduces new vulnerabilities. Adversaries can exploit retrieval mechanisms by injecting manipulated documents into the corpus~\cite{zhong-etal-2023-poisoning,clop2024backdooredretrieverspromptinjection,10.1145/3605764.3623985,10.1145/3689932.3694764}, subtly altering rankings to mislead LLM outputs. As shown in \cref{fig:adversarial_attack_example}, these adversarial documents mimic query-relevant patterns, making them difficult to detect while degrading the reliability of retrieval-based LLM systems. 
In real-world applications, LLMs are increasingly used in search engines to provide direct answers to user queries, a process known as Answer Engine Optimization (AEO)~\cite{seoaeo}. By leveraging retrieval-time manipulation techniques, attackers can craft adversarial content that not only ranks higher in search results but also steers the generated answers toward harmful or misleading content~\cite{riskofsearch,venkit2024searchenginesaiera}.

\begin{figure*}[!ht]
    \centering
    \includegraphics[width=\linewidth]{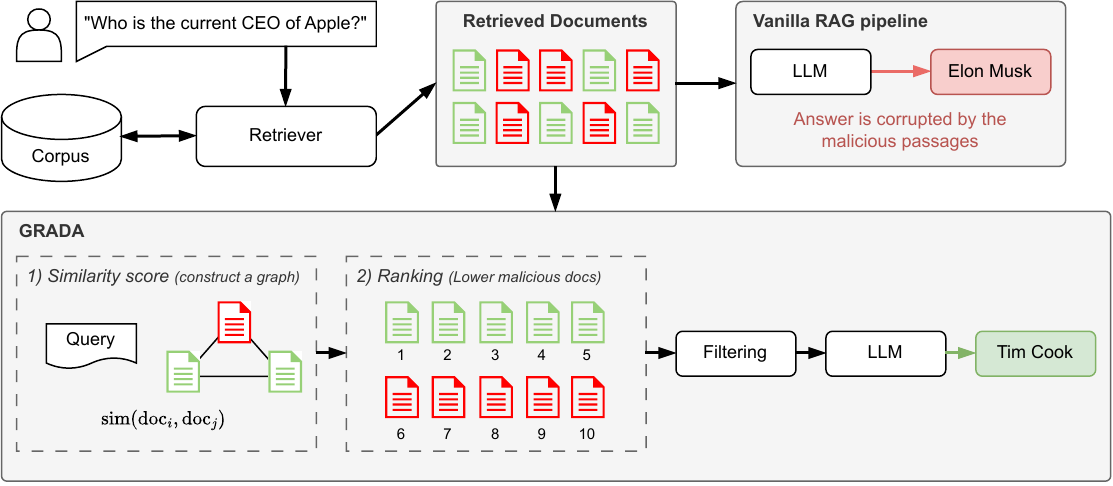}
    \caption{
        An overview of \ourmethod.
        A vanilla RAG pipeline concatenates all retrieved documents along with the question as input to the LLM.
        However, the accuracy of this pipeline can be easily harmed by malicious passages.
        In contrast, \ourmethod uses a graph-based approach to rerank and filter out malicious passages before passing the retrieved documents to LLMs for generation.
    }
    \label{fig:overview}
    \vspace{-2mm}
\end{figure*}

Existing noise filtering methods, such as Hybrid List Aware Transformer Reranking~\cite[HLATR,][]{Zhang2022HLATREM} and BAAI General Embeddings~\cite[BGE-reranker,][]{bge_embedding}, focus on improving document relevance by filtering out generic noise or low-quality content. However, these methods are ineffective against adversarial attacks that exploit query-document similarity patterns to evade detection. In addition, a recent study has reviewed current graph-based reranking methods \cite{zaoad2025graphbasedrerankingemergingtechniques}. It shows a potential path to use graphs in future information retrieval tasks, but the effects on adversarial documents remain unknown. On the other hand, specialized adversarial defenses, such as keyword filtering and decoding aggregation~\cite{xiang2024certifiablyrobustragretrieval}, can successfully remove adversarial content but at the cost of discarding valuable benign documents, ultimately weakening retrieval performance. This trade-off highlights the need for a more nuanced defense mechanism that can distinguish between adversarial and benign documents without compromising retrieval quality.

To address this challenge, we propose \textbf{G}raph-based \textbf{R}eranking against \textbf{A}dversarial \textbf{D}ocument \textbf{A}ttacks (\ourmethod), an effective defense framework designed to protect RAG systems from adversarial retrieval manipulations. Our key insight is that adversarial documents, while optimized for high query similarity, exhibit weaker semantic coherence with genuinely relevant documents in the retrieval set. Leveraging this property, we construct a graph where each retrieved document is represented as a node, and edges capture document-document similarity relationships. By propagating ranking scores through this graph structure, our approach prioritizes clusters of semantically consistent documents while suppressing adversarially crafted outliers. As illustrated in \cref{fig:overview}, our method significantly enhances the robustness of RAG-based LLMs, mitigating adversarial influences while preserving the integrity of benign retrieval results.

We conducted comprehensive experiments on Natural Questions (NQ), MS-MARCO, and HotpotQA across six different models. Our method has shown at least a 30\% decrease in reducing the Attack Success Rate (ASR), with improvements of up to 80\% across various adversarial attack strategies.

We summarize our contributions as follows:
\begin{itemize}
    \item We introduce \ourmethod, which constructs a weighted similarity graph among retrieved documents and iteratively propagates scores to mitigate the impacts of adversarial passages.

    \item We introduce a novel scoring function that simultaneously considers both query-document and document-document correlations, thereby improving robustness against adversarial attempts to mimic the query.

    \item We conducted comprehensive experiments on three distinct datasets, evaluating our method against four representative attack types. The results consistently demonstrate that \ourmethod outperforms existing defense baselines.

\end{itemize}

\section{Related Work}

Adversarial manipulation in IR has a long history. \citet{ilprints771} categorize web-spam strategies into content-based, link-based, and behavior-based attacks, while \citet{ntoulas2006detecting} use statistical features to detect spam content. \citet{castillo2010adversarial} survey a range of traditional attacks like cloaking and redirection, which expose fundamental weaknesses that persist in modern neural retrieval systems.

When RAG systems came out, Corpus poisoning attacks~\cite{zhong-etal-2023-poisoning} and third-party API attacks~\cite{zhao2024-api-attacks} show a new potential attack surface on LLMs.
%
%
Later, prompt injection attacks were introduced to bypass the retriever and affect the generator successfully~\cite{10.1145/3605764.3623985,10.1145/3689932.3694764}.
However, compared to the prior work, these methods are unstable in the retrieved adversarial passages.
\textit{While these attacks are based on modern LLM‑based retrieval, adversarial manipulation of information‑retrieval systems has a much longer history that is instructive for our setting.}

More recently, PoisonedRAG~\cite{zou2024poisonedrag} was proposed as a more stable attack.
It uses two passages concatenated together, with one of them appended to guarantee the retrieval of the adversarial passage and one to achieve the adversarial goal on the generator, which is to steer the LLM generating the answers anticipated by the attacker.
PoisonedRAG inspired many subsequent attacks.
Phantom~\cite{chaudhari2024phantomgeneraltriggerattacks}, which introduces a trigger to the question and achieves the adversarial goal only when the trigger is shown in the query.
Another type of Prompt Injection Attack~\cite[PIA,][]{clop2024backdooredretrieverspromptinjection} leverages the guaranteed retrieval mechanism in PoisonedRAG. Unlike typical misinformation attacks, this variant targets broader adversarial objectives beyond merely spreading false information.

A recent study proposed a defense mechanism that generates responses independently and produces an output based on the majority vote~\cite{xiang2024certifiablyrobustragretrieval}. 
While effective in some settings, this strategy defends only at the generator stage, which can compromise accuracy when multiple documents must be integrated. 
In contrast, several recent works intervene earlier in the RAG pipeline. TrustRAG~\cite{zhou2025trustragenhancingrobustnesstrustworthiness} employs clustering and LLM-based conflict resolution to filter poisoned documents before they influence the generator, substantially reducing attack success rates while preserving accuracy. 
Complementary to this, traceback approaches such as RAGForensics~\cite{10.1145/3696410.3714756} focus on identifying and removing the poisoned texts within the knowledge base itself, ensuring that subsequent retrieval yields only benign passages. 
Building on these insights, \ourmethod strengthens defenses at the reranking stage, preventing malicious content from reaching the generator without sacrificing the benefits of multi-document retrieval.

\begin{figure}[t]
    \centering
    \includegraphics[width=0.9\linewidth]{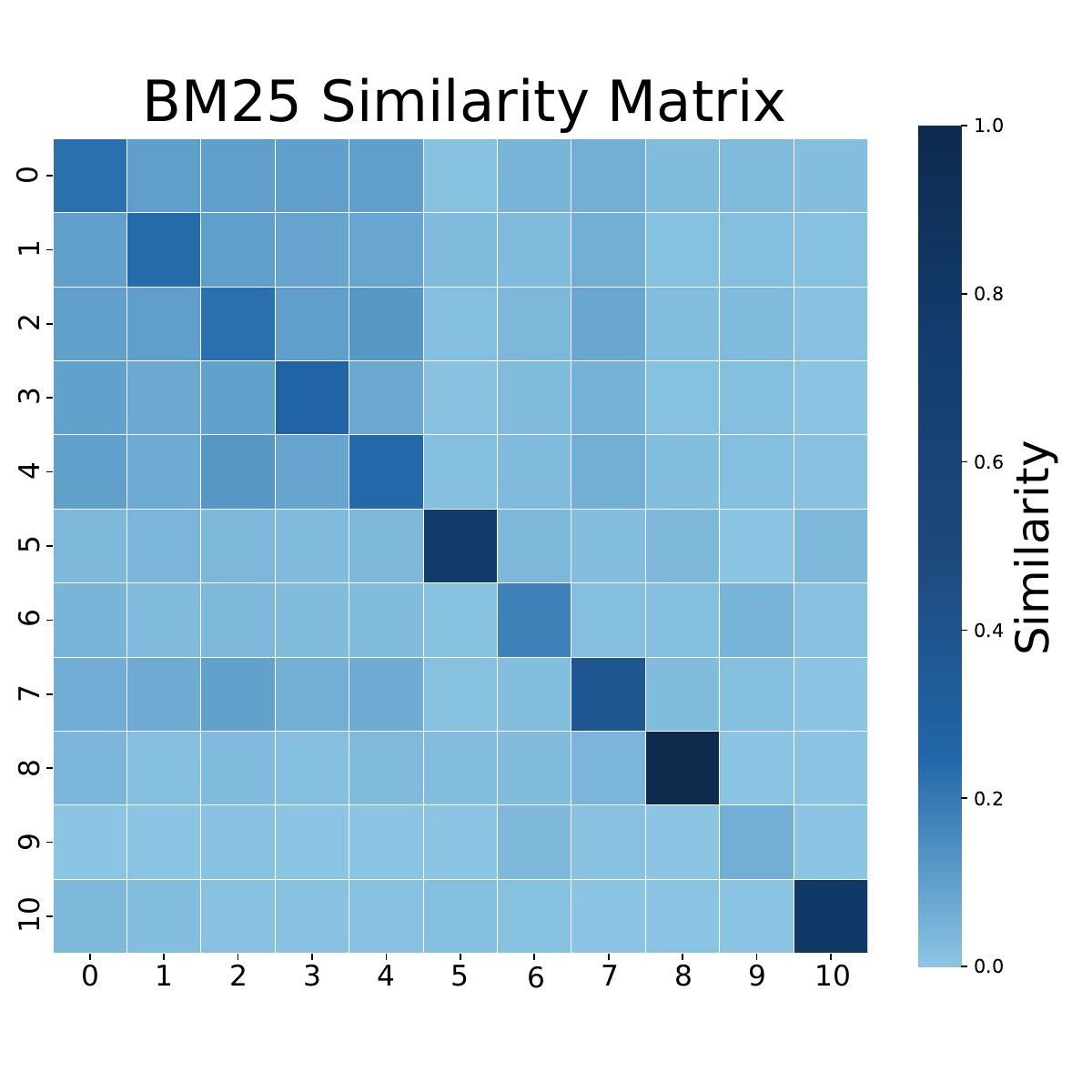}
    \caption{The similarity matrix among retrieved documents using BM25, where D0-D4 are poisoned and D5-D10 are clean documents.}
    \label{fig:Bm25_heatmap}
\end{figure}

\section{\ourmethod}
\label{sec:twostage}
A defining characteristic of recent poisoning attacks on RAG \cite{zou2024poisonedrag} is their focus on ensuring semantic similarity to the query while introducing anomalous similarity patterns among poisoned documents.
Adversarial documents closely resemble the query while diverging significantly from the legitimate documents, resulting in isolated patterns within the retrieval set, as illustrated in \cref{fig:overview,fig:Bm25_heatmap}.
Graph structures naturally capture these complex inter-document relationships by representing documents as nodes and similarities as edges.
Leveraging this intuition, we propose a graph-based reranking method that utilizes document-document similarity to enhance retrieval robustness.
In \cref{sec:graph}, we detail the graph construction process, followed by a description of our reranking system in \cref{sec:reranking}.

\subsection{Graph Construction}
\label{sec:graph}

We construct a weighted, undirected graph $\mathcal G = (V, E)$, where each node $v_i \in V$ corresponds to a document, and each edge $e_{ij} \in E$ is an undirected edge connecting node $v_i$ and $v_j$. Each edge is assigned a weight $w_{ij} \in [0, 1]$, which quantifies the similarity between the corresponding documents, \ie $\text{sim}(v_i, v_j)$. The graph is undirected because document relationships are not inherently directional; rather, the connectivity structure defines their associations.

The edge weight $w_{ij}$ can be computed using different approaches as follows:

\begin{itemize}[leftmargin=*]
    \item \textbf{Doc-to-Doc Similarity (D2DSIM):} The weight is directly determined by the similarity between documents.

\item \textbf{Hybrid Relevance Similarity (HRSIM):} A function $f$ that integrates both document-document similarity and query-document relevance:
\begin{equation*}
%
w_{ij} =  f\big(\text{sim}(v_i, v_j),\text{sim}(v_i, q),\text{sim}(v_j, q)\big).
\end{equation*}
\end{itemize}

The second approach assigns edge weights that not only reflect direct document-to-document similarity but also incorporate each document's relevance to an external query. This dual consideration leads to a more nuanced representation of document relationships.

To mitigate the influence of adversarial passages—documents that mimic the query $q$ to gain higher rankings—we introduce a function 
$f$, which adjusts the similarity score by applying a penalty based on the document-to-query similarities. First, we define the combined query relevance for a pair of documents $v_i$ and $v_j$ as follows:
\begin{equation*}
\text{sim}_{\text{sum}} = \text{sim}(v_i, q) + \text{sim}(v_j, q).
\end{equation*}
Then, the edge weight $w_{ij}$ between $v_i$ and $v_j$ is computed by subtracting a penalty term from their direct similarity, ensuring that the weight remains non-negative:
\begin{equation*}
w_{ij} = \max \{ \text{sim}(v_i, v_j) - \alpha \cdot \text{sim}_{\text{sum}}, 0 \}.
\end{equation*}
Here, $\alpha$ is a penalty coefficient that controls the influence of query similarity. If $\text{sim}(v_i, v_j) < \alpha \cdot \left[\text{sim}\left(v_i, q\right) + \text{sim}\left(v_i, q\right)\right]$, the edge weight is set to zero, which indicates removing the collusion connection between $v_i$ and $v_j$. 

Regarding the similarity function, we explore two popular methods:
\begin{itemize}[leftmargin=*]
 \item \textbf{BM25:} we use BM25~\cite{10.1561/1500000019} to calculate $\text{sim}(v_i, v_j)$. Since BM25 is an asymmetric metric, we adopt the following approach to compute the similarity score, ensuring symmetry in the process:
    \begin{equation*}
    w_{ij} = \frac{1}{2}\left(\text{BM25}\left(v_i, v_j\right) + \text{BM25}\left(v_j, v_i\right)\right).
    \end{equation*}
    
    \item \textbf{Embedding-based Distance (EBD):} we transform the documents $\mathbf{x}_i$ and $\mathbf{x}_j$ into dense vectors $v_i$ and $v_j$ and compute their cosine distance:
    \begin{equation*}
    w_{ij} = \text{sim}(v_i, v_j) = \frac{\mathbf{x}_i \cdot \mathbf{x}_j}{\|\mathbf{x}_i\| \|\mathbf{x}_j\|}.
    \end{equation*}
    
\end{itemize}

\subsection{Reranking}
\label{sec:reranking}

Inspired by PageRank~\cite{ilprints422}, we refine document rankings through an iterative score propagation process after constructing the graph. This approach prioritizes well-connected nodes while mitigating the influence of adversarial documents, ensuring a more robust and reliable ranking. 

Initially, each node $v_i$ is assigned a score $s^*_i$, forming the initial score vector $\mathbf{s}^* = [s^*_1, s^*_2, \ldots, s^*_n]^\top$.
The scores are then iteratively updated at each step $t$ via:
\begin{equation}
\begin{aligned}
    s_i^{(t)} &= (1 - d)s^*_j + d \sum_{v_j \in \mathcal{N}(i)} \frac{w_{ij}}{\sum_{v_k \in \mathcal{N}(j)} w_{jk}} s_j^{(t-1)},
\end{aligned}
\end{equation}
where $\mathcal N(i)$ represents the set of neighbor nodes connected by $v_i$ and $d$ is the damping factor, typically set to 0.85.\footnote{The experiments comparing different initialization methods are provided in \cref{app:initial_score_vector}.} The initial score vector $s^*$ is set by uniform initialization $s^* = \left[\frac{1}{|V|},\frac{1}{|V|},...,\frac{1}{|V|}\right]$.

The framework works as follows: The retriever identifies $M$ documents most similar to the query, with $n$ being the number of documents originally intended for retrieval.
To prevent adversarial documents from dominating the retrieved set, we ensure that poisoned documents do not constitute the majority by retrieving at least twice the number of documents (\ie $M \geq 2n$). For instance, if all $n$ original documents are poisoned (\eg $n=5$), incorporating at least $n$ additional benign documents guarantees that the majority of the final selection is non-poisoned. This approach maintains a substantial presence of benign content in the retrieved set, thereby improving the system’s resilience to adversarial manipulation.

After the algorithm reaches a stationary score distribution, the top $n$ documents are retained, while the remaining documents are discarded. Then, these top $n$ documents are provided as the context to the generative model.

\section{Experiments}
This section begins with the experimental setup, followed by a comparison of our approach with multiple baseline methods. Finally, we compare and analyze our approach across different settings.

\subsection{Experimental Setup}
\paragraph{Attack setup.} We conduct experiments on three widely used English datasets: \textbf{Natural Question}~\cite{kwiatkowski-etal-2019-natural}, \textbf{MS-MARCO}~\cite{DBLP:journals/corr/NguyenRSGTMD16} and \textbf{HotpotQA}~\cite{yang2018hotpotqa}. The victim models chosen for this study are \textbf{GPT-3.5-Turbo (version 0125)}~\cite{10.5555/3495724.3495883}, \textbf{GPT-4o (version 2024-08-06)}~\cite{openai2024gpt4technicalreport}, \textbf{Qwen2.5}~\cite{qwen2025qwen25technicalreport} and \textbf{LLaMA-3}~\cite{grattafiori2024llama3herdmodels}. The prompts used to generate answers are detailed in \cref{sec:appendix1}. Contriever~\cite{izacard2021contriever} is a dense retriever model used to find relevant documents by calculating similarity scores between the query and documents in the knowledge base. It was selected due to its efficiency and ability to handle large datasets. In this work, we investigate four distinct attack strategies on RAG. Two of them are Black-box attacks that have no knowledge about the retriever: PoisonedRAG~\cite{zou2024poisonedrag} and PIA~\cite{10.1145/3605764.3623985,10.1145/3689932.3694764,perez2022ignorepreviouspromptattack}. The remaining two are white-box attacks, in which attackers have the access to the victim's retriever, \ie PoisonedRAG~(Hotflip)~\cite{zou2024poisonedrag} and Phantom~\cite{chaudhari2024phantomgeneraltriggerattacks}

Under default settings without defense, as in~\citet{zou2024poisonedrag}, we retrieve the five most similar documents from the knowledge database to serve as the context for each question. We select 100 close-ended questions from each dataset, yielding 300 questions in total per attack-defense experiment. Additionally, this process is repeated using 3 random seeds, meaning each attack-defense pair is evaluated on 900 questions in total.

In contrast, \citet{zou2024poisonedrag} generated five poisoned texts and injected them into the retriever knowledge base.
To provide a more realistic assessment of the attack's effectiveness, we modify the experiment to inject only a single poisoned document into the database. The original setup, which retrieved only poisoned documents, resulted in a 100\% Attack Success Rate (ASR), making it impractical to evaluate the true impact of the attack. As shown in \cref{fig:Bm25_heatmap}, a similarity matrix appears to cluster poisoned documents in the top-left corner. By applying a clustering algorithm, we can identify and merge redundant information, effectively removing repetitive poisoned entries. This adjustment ensures that only one poisoned document is retrieved, allowing for a more meaningful evaluation of the attack performance.

\begin{table*}[t]
\centering
\resizebox{1.0\linewidth}{!}{%
\begin{tabular}{@{}lcccccc@{}}
\toprule
\multirow{3}{*}{\textbf{Defense}} &
\multicolumn{3}{c}{\textbf{PoisonedRAG}} &
\multicolumn{3}{c}{\textbf{PIA}} \\
\cmidrule(lr){2-4} \cmidrule(lr){5-7}  
 & \textbf{HotpotQA} & \textbf{NQ} & \textbf{MS-MARCO} & \textbf{HotpotQA} & \textbf{NQ} & \textbf{MS-MARCO}  \\
& \small{ASR $\downarrow$ / EM $\uparrow$} & \small{ASR $\downarrow$ / EM $\uparrow$} & \small{ASR $\downarrow$ / EM $\uparrow$} & \small{ASR $\downarrow$ / EM $\uparrow$} & \small{ASR $\downarrow$ / EM $\uparrow$} & \small{ASR $\downarrow$ / EM $\uparrow$}  \\ \midrule
 
\multicolumn{7}{c}{\textbf{\emph{GPT-3.5-Turbo}}}\\
\midrule
None               & 59.0±1.4 / 32.3±0.5 & 55.7±1.2 / 33.3±1.1& 46.5±1.5 / 41.0±0.0& 100.0±0.0 / 0.0±0.0& 98.0±0.0 / 2.0±0.0& 88.0±0.0 / 7.7±0.5\\
HLATR              & 62.3±0.5 / 30.3±0.5 & 51.5±0.5 / 35.5±0.5& 36.5±1.5 / 52.0±1.0& 100.0±0.0 / 0.0±0.0& 92.0±0.0 / 4.0±0.0& 84.0±0.0 / 9.0±0.0\\
BGE-reranker       & 56.6±0.9 / 36.3±1.2 & 46.5±0.5 / 43.5±0.5& 34.0±0.0 / 55.0±0.0& 98.0±0.0 / 2.0±0.0& 43.0±0.0 / 35.7±0.5& 43.0±0.0 / 43.0±0.8\\ 
Keyword Aggregation   & \cellcolor{cellgood}\textbf{11.0}±2.0 / 62.5±2.5 &  \cellcolor{cellgood}\textbf{2.0}±0.0 / 54.0±0.0& \cellcolor{cellgood}\textbf{3.0}±0.0 / 60.0±2.0& \cellcolor{cellgood}\textbf{0.0}±0.0 / 59.0±1.0&  \cellcolor{cellgood}\textbf{0.0}±0.0 / 48.0±0.0&  \cellcolor{cellgood}\textbf{0.0}±0.0 / 57.5±0.5\\
\ourmethod (D2DSIM-EBD)  & 48.6±1.2 / 39.0±0.8 & 26.1±1.0 / 50.9±1.0& 29.0±1.0 / 55.0±1.0& 33.0±0.0 / 42.3±0.5&  2.0±0.0 / 58.3±0.5&  3.0±0.0 / 70.5±0.5\\
\ourmethod (D2DSIM-BM25)    & 45.0±0.8 / 40.0±0.5 & 13.5±0.7 / 55.0±1.4& 16.5±0.5 / 65.5±0.5& 42.0±0.0 / 33.0±0.8&  12.0±0.0 / 55.3±0.5&  2.0±0.0 / 69.7±0.9\\
\ourmethod (HRSIM)  & \cellcolor{cellgood}\textbf{10.0}±0.0 / 51.0±0.8 &  \cellcolor{cellgood}\textbf{3.0}±0.6 / 58.0±1.1& \cellcolor{cellgood}\textbf{8.5}±0.5 / 71.5±0.5&  \cellcolor{cellgood}\textbf{27.0}±0.0 / 41.7±1.2&  \cellcolor{cellgood}\textbf{2.0}±0.0 / 61.7±2.1&  \cellcolor{cellgood}\textbf{1.0}±0.0 / 74.3±0.5\\
\midrule

\multicolumn{7}{c}{\textbf{\emph{Llama3.1-8b-Instruct}}}\\
\midrule
None                & 50.7±0.5 / 37.0±0.0& 49.0±0.8 / 33.0±0.8& 40.7±0.5 / 40.0±0.0& 88.3±0.5 / 3.0±0.0& 82.0±0.0 / 8.0±0.0& 69.0±0.0 / 14.0±0.0\\
HLATR              & 52.3±0.5 / 35.7±0.5&39.0±0.8 / 41.3±0.5& 35.7±0.5 / 43.3±0.5&91.3±0.5 / 2.7±0.5& 71.7±0.5 / 15.3±0.5& 50.0±0.8 / 19.7±0.5\\
BGE-reranker       & 51.7±0.5 / 36.0±0.0& 42.0±0.8 / 40.7±1.2& 33.7±0.5 / 42.0±0.8& 79.7±0.5 / 9.7±0.9& 30.0±0.0 / 40.3±0.5& 19.7±0.9 / 44.7±1.2 \\
Keyword Aggregation   & \cellcolor{cellgood}\textbf{6.7}±1.9 / 39.0±0.8& \cellcolor{cellgood}\textbf{3.0}±0.0 / 39.0±0.0& \cellcolor{cellgood}\textbf{6.7}±0.5 / 38.3±1.2& \cellcolor{cellgood}\textbf{0.0}±0.0 / 35.0±0.0& \cellcolor{cellgood}\textbf{0.0}±0.0 / 39.0±0.0& 0.0±0.0 / 36.0±0.8\\
\ourmethod (D2DSIM-EBD)  & 42.0±0.0 / 36.7±0.5& 24.0±0.0 / 46.7±0.5& 31.7±0.5 / 40.0±0.8& 30.7±0.5 / 35.3±0.90& \cellcolor{cellgood}\textbf{1.0}±0.0 / 55.3±0.5& 2.0±0.0 / 56.0±0.0\\
\ourmethod (D2DSIM-BM25)    & 30.0±0.0 / 39.3±0.5& 8.0±0.0 / 52.3±0.5& 19.3±0.5 / 49.7±0.9& 39.0±0.0 / 28.7±0.5& 7.7±0.5 / 48.3±0.9& \cellcolor{cellgood}\textbf{0.0}±0.0 / 55.0±0.0 \\
\ourmethod (HRSIM)  & \cellcolor{cellgood}\textbf{7.0}±0.0 / 44.0±0.8& \cellcolor{cellgood}\textbf{2.3}±0.5 / 55.7±0.5& \cellcolor{cellgood}\textbf{12.0}±0.0 / 52.3±0.5& \cellcolor{cellgood}\textbf{23.0}±0.0 / 36.7±0.5& 2.0±0.0 / 55.0±0.8& \cellcolor{cellgood}\textbf{0.0}±0.0 / 59.3±0.5\\
\bottomrule

\end{tabular}
}
\caption{ASR and EM (\%) for various defense methods on the black-box attack methods on GPT-3.5-Turbo and Llama3.1-8b-Instruct. The results of other models can be found in \cref{tab:other_poisoned,tab:benign_others,tab:other_hotflip,tab:other_phantom,tab:other_PIA}. We highlight the top-2 lowest ASR results in \colorbox{cellgood}{\textbf{blue}} cells.
}
\label{tab:publication}
\end{table*}

\paragraph{Defense setup.} We explore three similarity score combinations for \ourmethod: Embedding-based Distance, BM25, and Hybrid Relevance Similarity with BM25 as the similarity function.\footnote{We examine other similarity functions in \cref{sec:side_studies}}
Here, we utilize Contriever to encode both documents and queries, while for BM25, we adopt the implementation provided by ~\citet{bm25s}.
%
We compare \ourmethod against two reranking models and one defense method: HLATR~\cite{Zhang2022HLATREM}, which achieved first place in the MS-MARCO Passage Ranking Leaderboard, BGE-reranker~\cite{bge_embedding}, which achieves a high precision score in ranking tasks, and Keyword Aggregation~\cite{xiang2024certifiablyrobustragretrieval}, the only existing defense specifically designed for RAG-based adversarial attacks.

We evaluate the effectiveness of these defense methods by integrating them into our two-stage retrieval system described in \cref{sec:twostage}. We initially retrieve $M=10$ documents, which are then reranked using the aforementioned methods (except for Keyword Aggregation). The top five ranked documents are subsequently provided as the context for the model to answer the query. This ensures that, regardless of the defense configuration, the model always receives a fixed number of five context documents to respond to the question. For Keyword Aggregation, which does not perform reranking, the model directly generates the output based on the algorithm's keyword selections.

\paragraph{Evaluation metrics.} In our experiments, we employ Attack Success Rate (ASR) and Exact Match (EM) as metrics.
%
%
ASR is defined as the ratio of successful attacks to the total number of attacks conducted.
An attack is considered successful if the intended poisoned answer appears as a substring within the generated response from the model.
This definition accommodates attack strategies like PIA, which aim to introduce harmful links into the output of the model, allowing for some tolerance to semantically equivalent responses.
A higher ASR indicates a more successful attack.
This evaluation methodology follows the approach used in previous work~\cite{zou2024poisonedrag}.

To assess the question-answering accuracy of the models, we adopt EM score.
EM requires that the predicted answer of the model matches the ground truth answer exactly.
This strict criterion ensures that the response of the model is precise and follows the need for exact wording specified in the query, as outlined in \cref{sec:appendix1}.

\begin{table*}[t]
\centering
\resizebox{1.0\linewidth}{!}{%
\begin{tabular}{@{}lccccccc@{}}
\toprule
\multirow{3}{*}{\textbf{Defense}} &
\multicolumn{3}{c}{\textbf{PoisonedRAG(Hotflip)}} &
\multicolumn{3}{c}{\textbf{Phantom}} \\
\cmidrule(lr){2-4} \cmidrule(lr){5-7} 
 & \textbf{HotpotQA} & \textbf{NQ} & \textbf{MS-MARCO} & \textbf{HotpotQA} & \textbf{NQ} & \textbf{MS-MARCO} \\
& \small{ASR $\downarrow$ / EM $\uparrow$} & \small{ASR $\downarrow$ / EM $\uparrow$} & \small{ASR $\downarrow$ / EM $\uparrow$} & \small{ASR $\downarrow$ / EM $\uparrow$} & \small{ASR $\downarrow$ / EM $\uparrow$} & \small{ASR $\downarrow$ / EM $\uparrow$}  \\ \midrule
 
\multicolumn{7}{c}{\textbf{\emph{GPT-3.5-Turbo}}}\\
\midrule
None              & 62.0±0.8 / 29.3±0.5& 55.0±0.0 / 31.5±0.5& 42.5±0.5 / 47.5±0.5 & 99.0±0.0 / 1.0±0.0& 88.7±0.5 / 5.7±0.9& 67.7±1.9 / 25.7±1.7\\
HLATR             & 60.7±0.5 / 30.3±0.5& 49.6±0.9 / 36.0±0.8& 31.3±2.1 / 55.0±2.2 & 97.3±0.5 / 2.7±0.5& 90.7±0.5 / 7.0±0.8& 64.7±9.6 / 27.3±8.2\\
BGE-reranker      & 56.6±0.5 / 34.3±1.2& 43.0±0.8 / 40.7±0.5& 27.3±1.2 / 59.7±0.5 & 94.0±0.0 / 6.0±0.0& 70.7±4.7 / 17.3±0.5& 57.3±9.4 / 30.7±7.4\\ 
Keyword Aggregation  & \cellcolor{cellgood}\textbf{12.0}±0.8 / 62.3±2.1&  \cellcolor{cellgood}\textbf{2.0}±0.0 / 52.0±4.0&  \cellcolor{cellgood}\textbf{4.0}±0.8 / 57.0±2.6 &  \cellcolor{cellgood}\textbf{0.0}±0.0 / 50.0±0.8&  \cellcolor{cellgood}\textbf{0.0}±0.0 / 44.0±0.0&  \cellcolor{cellgood}\textbf{0.0}±0.0 / 57.0±1.0\\
\ourmethod (D2DSIM-EBD) & 44.7±0.9 / 39.3±1.2&  14.0±3.5 / 52.7±2.5&  10.7±1.2 / 69.0±0.0 & 60.7±0.5 / 19.7±0.5&  14.0±0.0 / 45.3±0.5&  13.0±0.0 / 59.0±2.2\\
\ourmethod (D2DSIM-BM25)  & 37.0±0.8 / 44.0±0.0&  9.0±0.0 / 59.3±0.5&  7.3±0.9 / 70.7±0.9 &  27.0±0.0 / 33.0±0.8&  5.7±0.5 / 50.0±0.8&  0.3±0.5 / 66.0±2.2\\
\ourmethod (HRSIM) & \cellcolor{cellgood}\textbf{7.3}±0.5 / 52.7±0.9&  \cellcolor{cellgood}\textbf{4.0}±0.0 / 58.3±1.2&  \cellcolor{cellgood}\textbf{6.3}±0.9 / 72.3±1.2 &  \cellcolor{cellgood}\textbf{23.0}±0.0 / 37.3±1.2&  \cellcolor{cellgood}\textbf{0.0}±0.0 / 48.5±0.5&  \cellcolor{cellgood}\textbf{0.0}±0.0 / 70.0±0.5\\
\midrule

\multicolumn{7}{c}{\textbf{\emph{Llama3.1-8b-Instruct}}}\\
\midrule
None               & 53.0±2.8 / 32.7±1.2& 50.0±1.4 / 30.0±2.2& 49.0±0.0 / 32.0±1.6 & 99.7±0.5 / 0.3±0.5& 89.3±2.1 / 9.3±1.2& 73.0±1.6 / 20.3±1.7\\
HLATR              & 53.3±2.9 / 32.7±2.1& 43.7±2.1 / 37.7±2.4& 36.0±1.4 / 37.7±1.7 & 96.7±1.2 / 3.0±0.8& 92.7±1.2 / 6.0±1.4& 72.3±1.2 / 18.0±1.6\\
BGE-reranker       & 50.0±3.7 / 34.3±0.5& 42.3±0.5 / 36.3±1.2& 27.3±1.2 / 59.7±0.5 & 95.3±1.2 / 3.0±0.8& 72.0±1.6 / 21.7±1.7& 62.0±0.8 / 26.0±1.6 \\
Keyword Aggregation  & \cellcolor{cellgood}\textbf{12.0}±0.8 / 62.3±2.1&  \cellcolor{cellgood}\textbf{2.0}±0.0 / 52.0±4.0&  \cellcolor{cellgood}\textbf{4.0}±0.8 / 57.0±2.6 & \cellcolor{cellgood}\textbf{0.0}±0.0 / 32.0±0.0& \cellcolor{cellgood}\textbf{0.0}±0.0 / 36.0±0.0& 0.0±0.0 / 39.7±0.9 \\
\ourmethod (D2DSIM-EBD) & 39.7±2.5 / 35.7±2.6& 13.0±0.0 / 50.7±2.1& 14.7±1.2 / 52.3±1.9 & 57.7±2.6 / 22.7±2.1& 10.7±1.9 / 48.7±1.2& 11.3±1.2 / 51.3±1.2\\
\ourmethod (D2DSIM-BM25)   & 32.0±0.8 / 38.0±0.0& 8.7±0.9 / 52.0±0.8& 13.3±0.9 / 53.0±0.8 & 26.7±0.5 / 37.0±2.2& 4.3±0.5 / 53.7±1.2& \cellcolor{cellgood}\textbf{1.0}±0.0 / 56.0±0.0 \\
\ourmethod (HRSIM) & \cellcolor{cellgood}\textbf{8.7}±1.7 / 44.7±1.2&  \cellcolor{cellgood}\textbf{2.0}±0.8 / 53.3±2.6&  \cellcolor{cellgood}\textbf{6.3}±0.9 / 72.3±1.2 & \cellcolor{cellgood}\textbf{10.3}±2.1 / 41.3±1.9& \cellcolor{cellgood}\textbf{0.3}±0.5 / 53.7±0.9& \cellcolor{cellgood}\textbf{0.0}±0.0 / 60.3±1.7\\
\bottomrule

\end{tabular}
}
\caption{Table 2: ASR and EM (\%) for various defense methods on the white-box attack methods on GPT-3.5-Turbo and Llama3.1-8b-Instruct.}
\label{tab:hotflip}
\vspace{-3mm}
\end{table*}

\begin{table}[t]
\centering
\resizebox{0.99\linewidth}{!}{%
\begin{tabular}{@{}lccc@{}}
\toprule
\textbf{Defense} & \textbf{HotpotQA} & \textbf{NQ} & \textbf{MS-MARCO} \\
\midrule
\multicolumn{4}{c}{\textbf{\emph{GPT-3.5-Turbo}}}\\
\midrule
No-RAG             & 16.3±1.7& 23.7±1.3& 11.7±0.5\\
None                           & 64.3±0.5& 58.6±1.2& 76.0±0.0\\
HLATR                          & \textbf{70.0}±0.8& 62.0±0.8& 77.7±0.5\\
BGE-reranker                   & 68.0±1.4& 64.7±1.2& \textbf{78.3}±0.5\\ 
Keyword Aggregation  & 68.3±0.5& 48.0±0.0& 59.0±0.0\\
\ourmethod (D2DSIM-EBD)            & 64.0±0.8& 61.0±0.8& 74.3±0.5\\
\ourmethod (D2DSIM-BM25)              & 57.3±0.5& \textbf{64.7}±0.5& 75.0±1.6\\
\ourmethod (HRSIM) & 50.0±0.5& 62.0±0.0& 75.3±0.5\\
\midrule 

\multicolumn{4}{c}{\textbf{\emph{Llama3.1-8b-Instruct}}}\\
\midrule
No-RAG             & 4.3±0.5& 3.0±0.0& 3.7±1.2\\
None               & 56.7±0.5& 50.0±0.0& 55.0±0.0\\
HLATR              & 56.0±0.8& 51.0±0.0& 56.3±0.5\\
BGE-reranker       & \textbf{58.0}±0.8& 54.0±0.8& \textbf{59.3}±0.9\\
Keyword Aggregation  & 34.0±0.0& 39.0±0.0& 36.0±0.8\\
\ourmethod (D2DSIM-EBD)  & 52.0±1.4& \textbf{54.7}±0.5&  58.3±0.5\\
\ourmethod (D2DSIM-BM25)    & 47.0±0.8& 52.7±0.5& 54.3±0.9\\
\ourmethod (HRSIM) & 43.3±0.9& \textbf{54.7}±0.5& 57.0±0.8\\
                               \bottomrule

\end{tabular}
}
\caption{EM scores of defense methods when presented with benign inputs. }
\label{tab:benign}
\vspace{-2mm}
\end{table}

\subsection{Results and Discussions}
\paragraph{Attacking without defense.} As shown in \cref{tab:publication}, including a single poisoned document in the retrieval process results in a high ASR. For instance, PoisonedRAG achieves an ASR of around 50\% across three datasets on both GPT-3.5-Turbo and  Llama3.1-8b-Instruct. PIA achieves at least 69\% ASR on Llama3.1-8b-Instruct and up to 100\% ASR in GPT-3.5-Turbo. These findings emphasize that even minimal adversarial input can achieve very high ASR and degrade the model's accuracy.

\paragraph{Effectiveness of \ourmethod.} The impact of \ourmethod on mitigating adversarial attacks is demonstrated in \cref{tab:publication,tab:hotflip}. As shown in \cref{tab:publication}, on the NQ and MS-MARCO datasets using GPT-3.5-Turbo, the ASR for PIA decreases from 98.0\% and 88.0\% to 2.0\% and 3.0\% by using D2DSIM-EBD. With D2DSIM-EBD, \ourmethod is also effective against PoisonedRAG, effectively reducing ASRs from 55.7\% and 46.5\% to 26.1\% and 29.0\%. However, the reduction of ASR against PoisonedRAG is more modest than against the other attacks. In this attack, D2DSIM-BM25 and HRSIM led to significant improvements compared to D2DSIM-EBD, where D2DSIM-BM25 achieved an extra 13\% decrease in ASR to 13.5\% and 16.5\%. Beyond that, HRSIM which introduces penalties for excessive similarity to the query, finalizes the ASR to 3\% and 8.5\%.


The defense methods demonstrate consistent effectiveness across the NQ and MS-MARCO datasets, achieving ASR reductions of over 30\% in most cases. However, performance on HotpotQA is less stable, particularly for D2DSIM-EBD and D2DSIM-BM25, which achieve only around a 10\% reduction in ASR against PoisonedRAG attacks. In contrast, HRSIM maintains its effectiveness, delivering ASR reductions exceeding 30\%, comparable to its performance on other datasets. This discrepancy likely stems from HotpotQA's multi-hop reasoning requirements, which pose challenges for single-document similarity metrics.

In \cref{tab:publication}, HLATR and BGE-reranker exhibit limited ability to filter poisoned documents, with ASR remaining largely unchanged compared to scenarios without any defense mechanisms. Although BGE-reranker occasionally outperforms HLATR, its overall performance remains inferior to \ourmethod in handling adversarial cases. This discrepancy underscores a critical limitation in contemporary reranking systems, which are primarily optimized for question relevance but insufficiently equipped to address adversarial attacks with high question relevance. 

Keyword Aggregation is able to reduce ASR significantly, especially for attacks like PIA and Phantom.
Keyword Aggregation works by extracting keywords from the answers of each passage to generate the final response, effectively neutralizing attack payloads designed to manipulate or deny answers, such as producing advertisements. While Keyword Aggregation reduces ASR effectively, its EM scores are usually lower than those of \ourmethod. For example, on Llama3.1-8b-Instruct in \cref{tab:publication}, \ourmethod's EM scores dominate Keyword Aggregation with at most 21\% difference, as some critical information may be lost during keyword extraction. This shows the ability of \ourmethod to perform well on normal answers even after mitigating adversarial contents. 

Similar results to those presented in \cref{tab:publication} can be observed in \cref{tab:hotflip} as well. Notably, \ourmethod combined with HRSIM consistently outperforms all other approaches, demonstrating that HRSIM is a strong similarity scoring function compared to the alternatives used in \ourmethod.


\cref{tab:benign} highlights the impact of different defense mechanisms on benign inputs. On GPT-3.5-Turbo, both HLATR and BGE-reranker demonstrate strong performance, outperforming \ourmethod and enhancing the model's overall accuracy. These reranking systems yield at least a 2\% improvement in EM scores, suggesting their effectiveness in mitigating noise unrelated to the posed questions.

\ourmethod with D2DSIM-EBD effectively preserves model performance on benign inputs across all datasets, with EM score deviations remaining within 3\%. Notably, the use of D2DSIM-BM25 leads to a 6\% improvement in EM scores on NQ, matching the performance of BGE-reranker, which achieves the highest EM overall. However, on HotpotQA, HRSIM resulted in a 14\% reduction in EM scores when handling benign inputs. While this trade-off is significant, it corresponds to HRSIM’s remarkable ASR reduction. Striking a balance between retrieval quality and defense robustness remains a crucial challenge for future research.

Keyword Aggregation has a much lower performance also in EM scores on benign input compared to \ourmethod. For example, in MS-MARCO, it results in 36\% compared to 57\% on Llama3.1-8b-Instruct and 59\% compared to 75.3\% on GPT-3.5-Turbo. Indeed showing the cost of discarding valuable information when facing benign documents.

Using \ourmethod, we demonstrate that it is possible to defend against the chosen attacks without compromising the model's overall performance. While reranking methods such as HLATR and BGE-reranker show promise in reducing noise, their limited effectiveness in countering adversarial attack noise highlights a critical gap in existing defenses. Similarly, Keyword Aggregation presents a valuable strategy for mitigating attack payloads but comes with significant trade-offs in EM scores.

\paragraph{Why \ourmethod works.}

For effective attacks, adversaries should steer the retriever to select the poisoned documents. To accomplish this, they typically craft these documents to closely resemble the query, exploiting the fact that most retrieval models prioritize query-document similarity. However, these adversarial documents often exhibit only weak similarity to the rest of the corpus, a property that makes them less likely to be flagged by defense mechanisms based on inter-document similarity comparisons.

\ourmethod leverages this insight by constructing a document similarity graph in which each document effectively ``votes'' for other documents with which it shares strong semantic similarity. Benign documents, which naturally cluster around shared content, tend to form densely connected subgraphs with high mutual similarity (\eg averaging $0.82$), thereby reinforcing each other. In contrast, poisoned documents—engineered to deceive—are typically more isolated, receiving fewer ``votes'' due to their low average similarity to genuine documents (\eg $0.35$). As a result, \ourmethod amplifies the collective influence of benign content while attenuating the impact of sparsely connected adversarial documents. A running example is provided in \cref{fig:initial_draw} in Appendix.

\begin{figure}[t]
    \centering
        \centering
        \includegraphics[width=\linewidth]{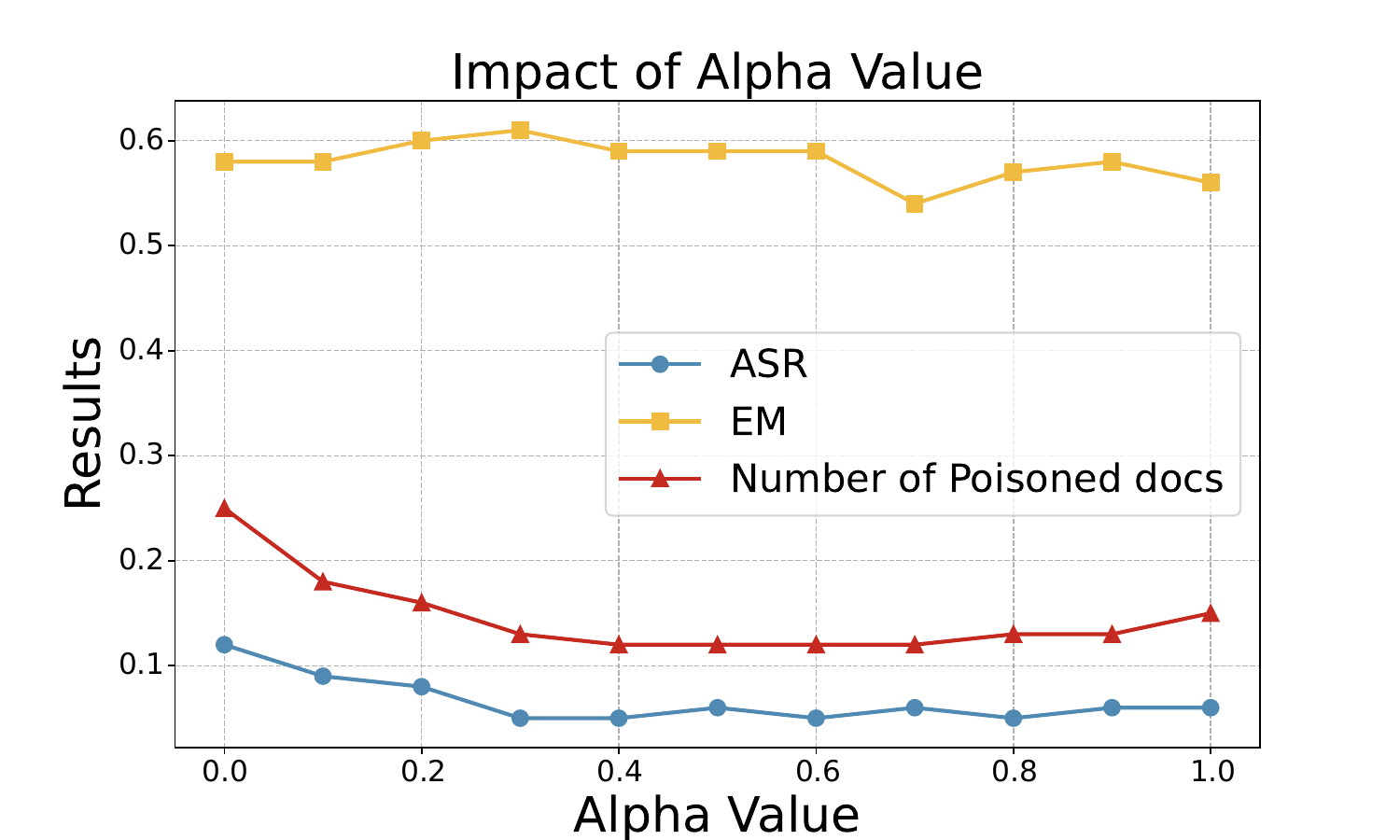}
        \caption{Comparison of $\alpha$ on three metrics (ASR, number of poisoned documents, and EM), based on NQ dataset with GPT-3.5-Turbo.}
        \label{fig:Impact_NQ}
\end{figure}

\paragraph{Impact of hyper-parameters $\alpha$ and $M$.} As shown in \cref{fig:Impact_NQ}, the number of poisoned documents in the context decreases as $\alpha$ increases, reaching a minimum at $\alpha = 0.3$ before starting to rise again after $\alpha = 0.8$. The ASR follows a similar trend to the number of poisoned documents after $\alpha = 0.3$. Conversely, the EM score exhibits a minimum at $\alpha = 0.7$. We selected $\alpha = 0.4$ because it strikes a balance, avoiding excessive penalization for query similarity, which could otherwise result in fewer query-related documents. When $\alpha = 0.4$, all three metrics (ASR, number of poisoned documents, and EM) are within an acceptable range, approaching the optimal performance values for $\alpha$.

    
\begin{figure}[t]
    \centering
        \includegraphics[width=\linewidth]{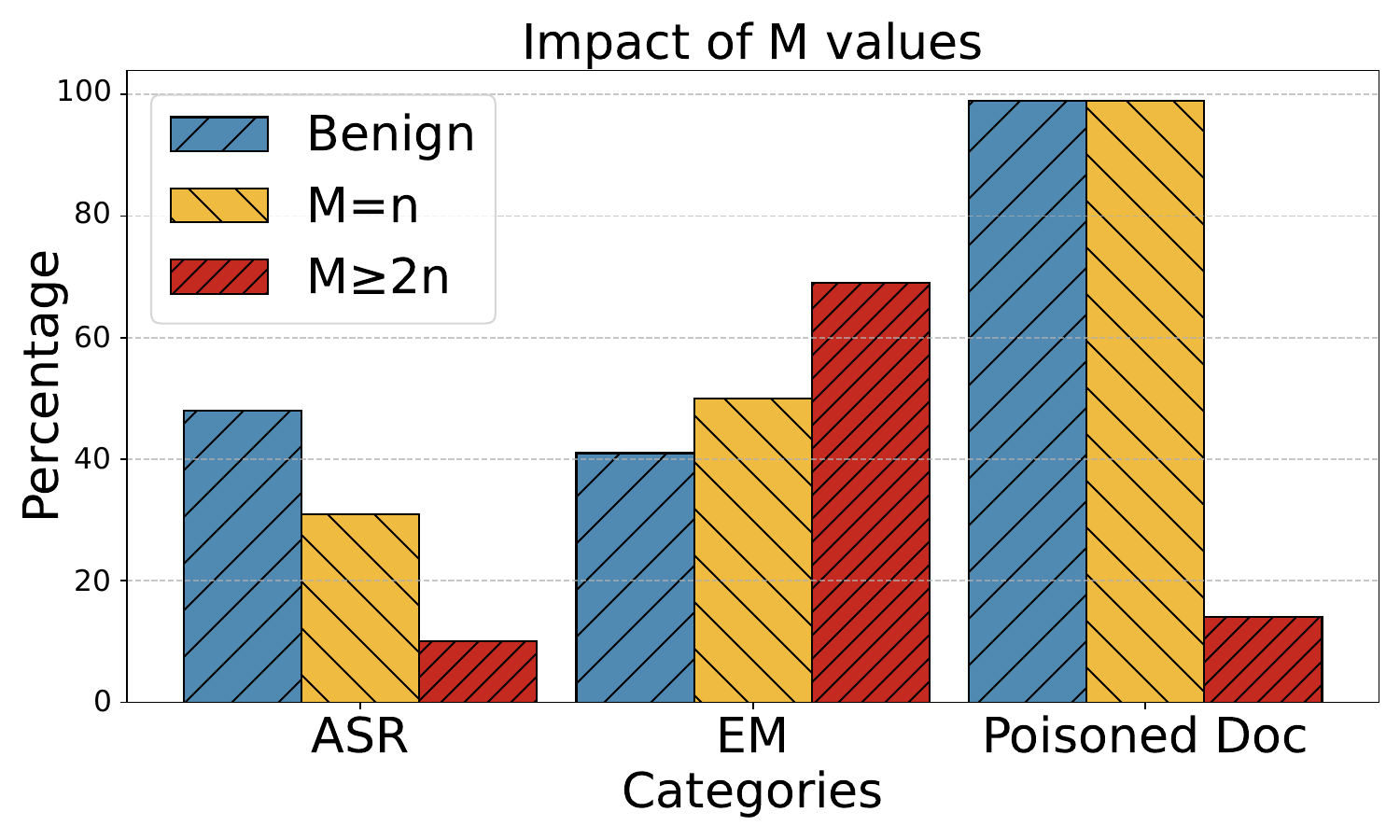}
        \caption{Comparison of $M$ value on three metrics (ASR, number of poisoned documents, and EM), based on
    MSMARCO dataset with GPT-3.5-Turbo.}
        \label{fig:Impact_M_MSMARCO}
\end{figure}

\cref{fig:Impact_M_MSMARCO} illustrates the effect of selecting $M = n$. It shows that, regardless of how documents are re-ranked, poisoned documents can still remain within the context provided to the model. However, this approach results in a 17\% decrease in ASR and a 9\% increase in EM, indicating that simply adjusting document positions can significantly impact model performance. This aligns with our observations in \cref{tab:pos_attack}, and the specific positions of the documents are detailed in \cref{fig:position_MSMARCO(BM25)}. By including additional documents for reranking and then retrieving only the top $n$ results, the ASR is further reduced from 21\% to 10\%, with only 14\% of poisoned documents remaining in the context provided to the model. This demonstrates the importance of including extra documents during reranking to remove poisoned content and achieve better overall performance effectively.

\section{Conclusion}
Our research examines the robustness challenges faced by RAG systems. We identify a critical vulnerability in current adversarial attacks, which focus on increasing semantic similarity to the query without accounting for the relationships between the retrieved documents. Our proposed graph-based filtering framework, \ourmethod, enhances the robustness of RAG systems by leveraging document similarities and effectively mitigating adversarial impacts through information flow. Experimental results on datasets such as MS-MARCO and NQ, demonstrate at least 30\% reductions in ASR across various adversarial strategies. Overall, this work presents a promising direction for developing more secure and reliable RAG systems.
\section*{Limitations}
Despite its effectiveness, our approach has limitations. First, it struggles with multi-hop reasoning tasks, facing attacks like PIA and Phantom. As the number of poisoned documents increases, system robustness deteriorates. Second, our method assumes poisoned documents are a minority. When they form the majority, their effectiveness declines, and future work should explore adaptive retrieval strategies to counter adversarial dominance.

\section*{Ethics Statement}

Our study focuses on improving the robustness of RAG systems, thereby enhancing their reliability and minimizing harmful manipulations.
We evaluated our proposed method, \ourmethod, using publicly available datasets as detailed in \cref{app:license}.
We do not engage in harmful data practices.

\section*{Acknowledgement}

Aryo Pradipta Gema was supported by the United Kingdom Research and Innovation (grant EP/S02431X/1), UKRI Centre for Doctoral Training in Biomedical AI at the University of Edinburgh, School of Informatics.
Giwon Hong was supported by the ILCC PhD program (School of Informatics Funding Package) at the University of Edinburgh, School of Informatics.
Xuanli He was funded by an industry grant from Cisco.
Pasquale Minervini was partially funded by ELIAI, an industry grant from Cisco, and a donation from Accenture LLP.
Qiongkai Xu acknowledges support from 2024 FSE Strategic Startup Grant.
This work was supported by the Edinburgh International Data Facility (EIDF) and the Data-Driven Innovation Programme at the University of Edinburgh.

\bibliography{custom}

\begin{thebibliography}{38}
\providecommand{\natexlab}[1]{#1}

\bibitem[{Brown et~al.(2020)Brown, Mann, Ryder, Subbiah, Kaplan, Dhariwal, Neelakantan, Shyam, Sastry, Askell, Agarwal, Herbert-Voss, Krueger, Henighan, Child, Ramesh, Ziegler, Wu, Winter, Hesse, Chen, Sigler, Litwin, Gray, Chess, Clark, Berner, McCandlish, Radford, Sutskever, and Amodei}]{10.5555/3495724.3495883}
Tom~B. Brown, Benjamin Mann, Nick Ryder, Melanie Subbiah, Jared Kaplan, Prafulla Dhariwal, Arvind Neelakantan, Pranav Shyam, Girish Sastry, Amanda Askell, Sandhini Agarwal, Ariel Herbert-Voss, Gretchen Krueger, Tom Henighan, Rewon Child, Aditya Ramesh, Daniel~M. Ziegler, Jeffrey Wu, Clemens Winter, and 12 others. 2020.
\newblock Language models are few-shot learners.
\newblock In \emph{Proceedings of the 34th International Conference on Neural Information Processing Systems}, NIPS '20, Red Hook, NY, USA. Curran Associates Inc.

\bibitem[{Castillo and Davison(2011)}]{castillo2010adversarial}
Carlos Castillo and Brian~D. Davison. 2011.
\newblock \href {https://doi.org/10.1561/1500000021} {\emph{Adversarial Web Search}}.
\newblock Now Foundations and Trends.

\bibitem[{Chaudhari et~al.(2024)Chaudhari, Severi, Abascal, Jagielski, Choquette-Choo, Nasr, Nita-Rotaru, and Oprea}]{chaudhari2024phantomgeneraltriggerattacks}
Harsh Chaudhari, Giorgio Severi, John Abascal, Matthew Jagielski, Christopher~A. Choquette-Choo, Milad Nasr, Cristina Nita-Rotaru, and Alina Oprea. 2024.
\newblock \href {https://arxiv.org/abs/2405.20485} {Phantom: General trigger attacks on retrieval augmented language generation}.
\newblock \emph{Preprint}, arXiv:2405.20485.

\bibitem[{Clop and Teglia(2024)}]{clop2024backdooredretrieverspromptinjection}
Cody Clop and Yannick Teglia. 2024.
\newblock \href {https://arxiv.org/abs/2410.14479} {Backdoored retrievers for prompt injection attacks on retrieval augmented generation of large language models}.
\newblock \emph{Preprint}, arXiv:2410.14479.

\bibitem[{Fourrier et~al.(2024)Fourrier, Habib, Lozovskaya, Szafer, and Wolf}]{open-llm-leaderboard-v2}
Clémentine Fourrier, Nathan Habib, Alina Lozovskaya, Konrad Szafer, and Thomas Wolf. 2024.
\newblock Open llm leaderboard v2.
\newblock \url{https://huggingface.co/spaces/open-llm-leaderboard/open_llm_leaderboard}.

\bibitem[{Graff et~al.(2003)Graff, Kong, Chen, and Maeda}]{graff2003english}
David Graff, Junbo Kong, Ke~Chen, and Kazuaki Maeda. 2003.
\newblock English gigaword.
\newblock \emph{Linguistic Data Consortium, Philadelphia}, 4(1):34.

\bibitem[{Grattafiori et~al.(2024)Grattafiori, Dubey, Jauhri, Pandey, Kadian, Al-Dahle, Letman, Mathur, Schelten, Vaughan, Yang, Fan, Goyal, Hartshorn, Yang, Mitra, Sravankumar, Korenev, Hinsvark, Rao, Zhang, Rodriguez, Gregerson, Spataru, Roziere, Biron, Tang, Chern, Caucheteux, Nayak, Bi, Marra, McConnell, Keller, Touret, Wu, Wong, Ferrer, Nikolaidis, Allonsius, Song, Pintz, Livshits, Wyatt, Esiobu, Choudhary, Mahajan, Garcia-Olano, Perino, Hupkes, Lakomkin, AlBadawy, Lobanova, Dinan, Smith, Radenovic, Guzmán, Zhang, Synnaeve, Lee, Anderson, Thattai, Nail, Mialon, Pang, Cucurell, Nguyen, Korevaar, Xu, Touvron, Zarov, Ibarra, Kloumann, Misra, Evtimov, Zhang, Copet, Lee, Geffert, Vranes, Park, Mahadeokar, Shah, van~der Linde, Billock, Hong, Lee, Fu, Chi, Huang, Liu, Wang, Yu, Bitton, Spisak, Park, Rocca, Johnstun, Saxe, Jia, Alwala, Prasad, Upasani, Plawiak, Li, Heafield, Stone, El-Arini, Iyer, Malik, Chiu, Bhalla, Lakhotia, Rantala-Yeary, van~der Maaten, Chen, Tan, Jenkins, Martin, Madaan, Malo, Blecher,
  Landzaat, de~Oliveira, Muzzi, Pasupuleti, Singh, Paluri, Kardas, Tsimpoukelli, Oldham, Rita, Pavlova, Kambadur, Lewis, Si, Singh, Hassan, Goyal, Torabi, Bashlykov, Bogoychev, Chatterji, Zhang, Duchenne, Çelebi, Alrassy, Zhang, Li, Vasic, Weng, Bhargava, Dubal, Krishnan, Koura, Xu, He, Dong, Srinivasan, Ganapathy, Calderer, Cabral, Stojnic, Raileanu, Maheswari, Girdhar, Patel, Sauvestre, Polidoro, Sumbaly, Taylor, Silva, Hou, Wang, Hosseini, Chennabasappa, Singh, Bell, Kim, Edunov, Nie, Narang, Raparthy, Shen, Wan, Bhosale, Zhang, Vandenhende, Batra, Whitman, Sootla, Collot, Gururangan, Borodinsky, Herman, Fowler, Sheasha, Georgiou, Scialom, Speckbacher, Mihaylov, Xiao, Karn, Goswami, Gupta, Ramanathan, Kerkez, Gonguet, Do, Vogeti, Albiero, Petrovic, Chu, Xiong, Fu, Meers, Martinet, Wang, Wang, Tan, Xia, Xie, Jia, Wang, Goldschlag, Gaur, Babaei, Wen, Song, Zhang, Li, Mao, Coudert, Yan, Chen, Papakipos, Singh, Srivastava, Jain, Kelsey, Shajnfeld, Gangidi, Victoria, Goldstand, Menon, Sharma, Boesenberg,
  Baevski, Feinstein, Kallet, Sangani, Teo, Yunus, Lupu, Alvarado, Caples, Gu, Ho, Poulton, Ryan, Ramchandani, Dong, Franco, Goyal, Saraf, Chowdhury, Gabriel, Bharambe, Eisenman, Yazdan, James, Maurer, Leonhardi, Huang, Loyd, Paola, Paranjape, Liu, Wu, Ni, Hancock, Wasti, Spence, Stojkovic, Gamido, Montalvo, Parker, Burton, Mejia, Liu, Wang, Kim, Zhou, Hu, Chu, Cai, Tindal, Feichtenhofer, Gao, Civin, Beaty, Kreymer, Li, Adkins, Xu, Testuggine, David, Parikh, Liskovich, Foss, Wang, Le, Holland, Dowling, Jamil, Montgomery, Presani, Hahn, Wood, Le, Brinkman, Arcaute, Dunbar, Smothers, Sun, Kreuk, Tian, Kokkinos, Ozgenel, Caggioni, Kanayet, Seide, Florez, Schwarz, Badeer, Swee, Halpern, Herman, Sizov, Guangyi, Zhang, Lakshminarayanan, Inan, Shojanazeri, Zou, Wang, Zha, Habeeb, Rudolph, Suk, Aspegren, Goldman, Zhan, Damlaj, Molybog, Tufanov, Leontiadis, Veliche, Gat, Weissman, Geboski, Kohli, Lam, Asher, Gaya, Marcus, Tang, Chan, Zhen, Reizenstein, Teboul, Zhong, Jin, Yang, Cummings, Carvill, Shepard, McPhie,
  Torres, Ginsburg, Wang, Wu, U, Saxena, Khandelwal, Zand, Matosich, Veeraraghavan, Michelena, Li, Jagadeesh, Huang, Chawla, Huang, Chen, Garg, A, Silva, Bell, Zhang, Guo, Yu, Moshkovich, Wehrstedt, Khabsa, Avalani, Bhatt, Mankus, Hasson, Lennie, Reso, Groshev, Naumov, Lathi, Keneally, Liu, Seltzer, Valko, Restrepo, Patel, Vyatskov, Samvelyan, Clark, Macey, Wang, Hermoso, Metanat, Rastegari, Bansal, Santhanam, Parks, White, Bawa, Singhal, Egebo, Usunier, Mehta, Laptev, Dong, Cheng, Chernoguz, Hart, Salpekar, Kalinli, Kent, Parekh, Saab, Balaji, Rittner, Bontrager, Roux, Dollar, Zvyagina, Ratanchandani, Yuvraj, Liang, Alao, Rodriguez, Ayub, Murthy, Nayani, Mitra, Parthasarathy, Li, Hogan, Battey, Wang, Howes, Rinott, Mehta, Siby, Bondu, Datta, Chugh, Hunt, Dhillon, Sidorov, Pan, Mahajan, Verma, Yamamoto, Ramaswamy, Lindsay, Lindsay, Feng, Lin, Zha, Patil, Shankar, Zhang, Zhang, Wang, Agarwal, Sajuyigbe, Chintala, Max, Chen, Kehoe, Satterfield, Govindaprasad, Gupta, Deng, Cho, Virk, Subramanian, Choudhury,
  Goldman, Remez, Glaser, Best, Koehler, Robinson, Li, Zhang, Matthews, Chou, Shaked, Vontimitta, Ajayi, Montanez, Mohan, Kumar, Mangla, Ionescu, Poenaru, Mihailescu, Ivanov, Li, Wang, Jiang, Bouaziz, Constable, Tang, Wu, Wang, Wu, Gao, Kleinman, Chen, Hu, Jia, Qi, Li, Zhang, Zhang, Adi, Nam, Yu, Wang, Zhao, Hao, Qian, Li, He, Rait, DeVito, Rosnbrick, Wen, Yang, Zhao, and Ma}]{grattafiori2024llama3herdmodels}
Aaron Grattafiori, Abhimanyu Dubey, Abhinav Jauhri, Abhinav Pandey, Abhishek Kadian, Ahmad Al-Dahle, Aiesha Letman, Akhil Mathur, Alan Schelten, Alex Vaughan, Amy Yang, Angela Fan, Anirudh Goyal, Anthony Hartshorn, Aobo Yang, Archi Mitra, Archie Sravankumar, Artem Korenev, Arthur Hinsvark, and 542 others. 2024.
\newblock \href {https://arxiv.org/abs/2407.21783} {The llama 3 herd of models}.
\newblock \emph{Preprint}, arXiv:2407.21783.

\bibitem[{Greshake et~al.(2023)Greshake, Abdelnabi, Mishra, Endres, Holz, and Fritz}]{10.1145/3605764.3623985}
Kai Greshake, Sahar Abdelnabi, Shailesh Mishra, Christoph Endres, Thorsten Holz, and Mario Fritz. 2023.
\newblock \href {https://doi.org/10.1145/3605764.3623985} {Not what you've signed up for: Compromising real-world llm-integrated applications with indirect prompt injection}.
\newblock In \emph{Proceedings of the 16th ACM Workshop on Artificial Intelligence and Security}, AISec '23, page 79–90, New York, NY, USA. Association for Computing Machinery.

\bibitem[{Gyongyi and Garcia-Molina(2005)}]{ilprints771}
Zoltan Gyongyi and Hector Garcia-Molina. 2005.
\newblock \href {http://ilpubs.stanford.edu:8090/771/} {Web spam taxonomy}.
\newblock In \emph{First International Workshop on Adversarial Information Retrieval on the Web (AIRWeb 2005)}.

\bibitem[{Hammond(2024)}]{riskofsearch}
Kristian Hammond. 2024.
\newblock \href {https://casmi.northwestern.edu/news/articles/2024/the-risk-of-googles-shift-from-search-engine-to-answer-machine.html} {The risk of google’s shift from search engine to answer machine}.
\newblock \emph{CENTER FOR ADVANCING SAFETY OF MACHINE INTELLIGENCE}.

\bibitem[{Izacard et~al.(2021)Izacard, Caron, Hosseini, Riedel, Bojanowski, Joulin, and Grave}]{izacard2021contriever}
Gautier Izacard, Mathilde Caron, Lucas Hosseini, Sebastian Riedel, Piotr Bojanowski, Armand Joulin, and Edouard Grave. 2021.
\newblock \href {https://doi.org/10.48550/ARXIV.2112.09118} {Unsupervised dense information retrieval with contrastive learning}.

\bibitem[{Kwiatkowski et~al.(2019)Kwiatkowski, Palomaki, Redfield, Collins, Parikh, Alberti, Epstein, Polosukhin, Devlin, Lee, Toutanova, Jones, Kelcey, Chang, Dai, Uszkoreit, Le, and Petrov}]{kwiatkowski-etal-2019-natural}
Tom Kwiatkowski, Jennimaria Palomaki, Olivia Redfield, Michael Collins, Ankur Parikh, Chris Alberti, Danielle Epstein, Illia Polosukhin, Jacob Devlin, Kenton Lee, Kristina Toutanova, Llion Jones, Matthew Kelcey, Ming-Wei Chang, Andrew~M. Dai, Jakob Uszkoreit, Quoc Le, and Slav Petrov. 2019.
\newblock \href {https://doi.org/10.1162/tacl_a_00276} {Natural questions: A benchmark for question answering research}.
\newblock \emph{Transactions of the Association for Computational Linguistics}, 7:452--466.

\bibitem[{Lewis et~al.(2020)Lewis, Perez, Piktus, Petroni, Karpukhin, Goyal, K\"{u}ttler, Lewis, Yih, Rockt\"{a}schel, Riedel, and Kiela}]{10.5555/3495724.3496517}
Patrick Lewis, Ethan Perez, Aleksandra Piktus, Fabio Petroni, Vladimir Karpukhin, Naman Goyal, Heinrich K\"{u}ttler, Mike Lewis, Wen-tau Yih, Tim Rockt\"{a}schel, Sebastian Riedel, and Douwe Kiela. 2020.
\newblock Retrieval-augmented generation for knowledge-intensive nlp tasks.
\newblock In \emph{Proceedings of the 34th International Conference on Neural Information Processing Systems}, NIPS '20, Red Hook, NY, USA. Curran Associates Inc.

\bibitem[{Liu et~al.(2024)Liu, Lin, Hewitt, Paranjape, Bevilacqua, Petroni, and Liang}]{liu-etal-2024-lost}
Nelson~F. Liu, Kevin Lin, John Hewitt, Ashwin Paranjape, Michele Bevilacqua, Fabio Petroni, and Percy Liang. 2024.
\newblock \href {https://doi.org/10.1162/tacl_a_00638} {Lost in the middle: How language models use long contexts}.
\newblock \emph{Transactions of the Association for Computational Linguistics}, 12:157--173.

\bibitem[{Lù(2024)}]{bm25s}
Xing~Han Lù. 2024.
\newblock \href {https://arxiv.org/abs/2407.03618} {Bm25s: Orders of magnitude faster lexical search via eager sparse scoring}.
\newblock \emph{Preprint}, arXiv:2407.03618.

\bibitem[{Nguyen et~al.(2016)Nguyen, Rosenberg, Song, Gao, Tiwary, Majumder, and Deng}]{DBLP:journals/corr/NguyenRSGTMD16}
Tri Nguyen, Mir Rosenberg, Xia Song, Jianfeng Gao, Saurabh Tiwary, Rangan Majumder, and Li~Deng. 2016.
\newblock \href {https://arxiv.org/abs/1611.09268} {{MS} {MARCO:} {A} human generated machine reading comprehension dataset}.
\newblock \emph{CoRR}, abs/1611.09268.

\bibitem[{Ntoulas et~al.(2006)Ntoulas, Najork, Manasse, and Fetterly}]{ntoulas2006detecting}
Alexandros Ntoulas, Marc Najork, Mark Manasse, and Dennis Fetterly. 2006.
\newblock \href {https://doi.org/10.1145/1135777.1135794} {Detecting spam web pages through content analysis}.
\newblock In \emph{Proceedings of the 15th International Conference on World Wide Web}, WWW '06, page 83–92, New York, NY, USA. Association for Computing Machinery.

\bibitem[{OpenAI et~al.(2024)OpenAI, Achiam, Adler, Agarwal, Ahmad, Akkaya, Aleman, Almeida, Altenschmidt, Altman, Anadkat, Avila, Babuschkin, Balaji, Balcom, Baltescu, Bao, Bavarian, Belgum, Bello, Berdine, Bernadett-Shapiro, Berner, Bogdonoff, Boiko, Boyd, Brakman, Brockman, Brooks, Brundage, Button, Cai, Campbell, Cann, Carey, Carlson, Carmichael, Chan, Chang, Chantzis, Chen, Chen, Chen, Chen, Chen, Chess, Cho, Chu, Chung, Cummings, Currier, Dai, Decareaux, Degry, Deutsch, Deville, Dhar, Dohan, Dowling, Dunning, Ecoffet, Eleti, Eloundou, Farhi, Fedus, Felix, Fishman, Forte, Fulford, Gao, Georges, Gibson, Goel, Gogineni, Goh, Gontijo-Lopes, Gordon, Grafstein, Gray, Greene, Gross, Gu, Guo, Hallacy, Han, Harris, He, Heaton, Heidecke, Hesse, Hickey, Hickey, Hoeschele, Houghton, Hsu, Hu, Hu, Huizinga, Jain, Jain, Jang, Jiang, Jiang, Jin, Jin, Jomoto, Jonn, Jun, Kaftan, Łukasz Kaiser, Kamali, Kanitscheider, Keskar, Khan, Kilpatrick, Kim, Kim, Kim, Kirchner, Kiros, Knight, Kokotajlo, Łukasz Kondraciuk,
  Kondrich, Konstantinidis, Kosic, Krueger, Kuo, Lampe, Lan, Lee, Leike, Leung, Levy, Li, Lim, Lin, Lin, Litwin, Lopez, Lowe, Lue, Makanju, Malfacini, Manning, Markov, Markovski, Martin, Mayer, Mayne, McGrew, McKinney, McLeavey, McMillan, McNeil, Medina, Mehta, Menick, Metz, Mishchenko, Mishkin, Monaco, Morikawa, Mossing, Mu, Murati, Murk, Mély, Nair, Nakano, Nayak, Neelakantan, Ngo, Noh, Ouyang, O'Keefe, Pachocki, Paino, Palermo, Pantuliano, Parascandolo, Parish, Parparita, Passos, Pavlov, Peng, Perelman, de~Avila Belbute~Peres, Petrov, de~Oliveira~Pinto, Michael, Pokorny, Pokrass, Pong, Powell, Power, Power, Proehl, Puri, Radford, Rae, Ramesh, Raymond, Real, Rimbach, Ross, Rotsted, Roussez, Ryder, Saltarelli, Sanders, Santurkar, Sastry, Schmidt, Schnurr, Schulman, Selsam, Sheppard, Sherbakov, Shieh, Shoker, Shyam, Sidor, Sigler, Simens, Sitkin, Slama, Sohl, Sokolowsky, Song, Staudacher, Such, Summers, Sutskever, Tang, Tezak, Thompson, Tillet, Tootoonchian, Tseng, Tuggle, Turley, Tworek, Uribe, Vallone,
  Vijayvergiya, Voss, Wainwright, Wang, Wang, Wang, Ward, Wei, Weinmann, Welihinda, Welinder, Weng, Weng, Wiethoff, Willner, Winter, Wolrich, Wong, Workman, Wu, Wu, Wu, Xiao, Xu, Yoo, Yu, Yuan, Zaremba, Zellers, Zhang, Zhang, Zhao, Zheng, Zhuang, Zhuk, and Zoph}]{openai2024gpt4technicalreport}
OpenAI, Josh Achiam, Steven Adler, Sandhini Agarwal, Lama Ahmad, Ilge Akkaya, Florencia~Leoni Aleman, Diogo Almeida, Janko Altenschmidt, Sam Altman, Shyamal Anadkat, Red Avila, Igor Babuschkin, Suchir Balaji, Valerie Balcom, Paul Baltescu, Haiming Bao, Mohammad Bavarian, Jeff Belgum, and 262 others. 2024.
\newblock \href {https://arxiv.org/abs/2303.08774} {Gpt-4 technical report}.
\newblock \emph{Preprint}, arXiv:2303.08774.

\bibitem[{Page et~al.(1999)Page, Brin, Motwani, and Winograd}]{ilprints422}
Lawrence Page, Sergey Brin, Rajeev Motwani, and Terry Winograd. 1999.
\newblock \href {http://ilpubs.stanford.edu:8090/422/} {The pagerank citation ranking: Bringing order to the web.}
\newblock Technical Report 1999-66, Stanford InfoLab.
\newblock Previous number = SIDL-WP-1999-0120.

\bibitem[{Pasquini et~al.(2024)Pasquini, Strohmeier, and Troncoso}]{10.1145/3689932.3694764}
Dario Pasquini, Martin Strohmeier, and Carmela Troncoso. 2024.
\newblock \href {https://doi.org/10.1145/3689932.3694764} {Neural exec: Learning (and learning from) execution triggers for prompt injection attacks}.
\newblock In \emph{Proceedings of the 2024 Workshop on Artificial Intelligence and Security}, AISec '24, page 89–100, New York, NY, USA. Association for Computing Machinery.

\bibitem[{Perez and Ribeiro(2022)}]{perez2022ignorepreviouspromptattack}
Fábio Perez and Ian Ribeiro. 2022.
\newblock \href {https://arxiv.org/abs/2211.09527} {Ignore previous prompt: Attack techniques for language models}.
\newblock \emph{Preprint}, arXiv:2211.09527.

\bibitem[{Qwen et~al.(2025)Qwen, :, Yang, Yang, Zhang, Hui, Zheng, Yu, Li, Liu, Huang, Wei, Lin, Yang, Tu, Zhang, Yang, Yang, Zhou, Lin, Dang, Lu, Bao, Yang, Yu, Li, Xue, Zhang, Zhu, Men, Lin, Li, Tang, Xia, Ren, Ren, Fan, Su, Zhang, Wan, Liu, Cui, Zhang, and Qiu}]{qwen2025qwen25technicalreport}
Qwen, :, An~Yang, Baosong Yang, Beichen Zhang, Binyuan Hui, Bo~Zheng, Bowen Yu, Chengyuan Li, Dayiheng Liu, Fei Huang, Haoran Wei, Huan Lin, Jian Yang, Jianhong Tu, Jianwei Zhang, Jianxin Yang, Jiaxi Yang, Jingren Zhou, and 25 others. 2025.
\newblock \href {https://arxiv.org/abs/2412.15115} {Qwen2.5 technical report}.
\newblock \emph{Preprint}, arXiv:2412.15115.

\bibitem[{Reimers and Gurevych(2019)}]{reimers-gurevych-2019-sentence}
Nils Reimers and Iryna Gurevych. 2019.
\newblock \href {https://doi.org/10.18653/v1/D19-1410} {Sentence-{BERT}: Sentence embeddings using {S}iamese {BERT}-networks}.
\newblock In \emph{Proceedings of the 2019 Conference on Empirical Methods in Natural Language Processing and the 9th International Joint Conference on Natural Language Processing (EMNLP-IJCNLP)}, pages 3982--3992, Hong Kong, China. Association for Computational Linguistics.

\bibitem[{Robertson and Zaragoza(2009)}]{10.1561/1500000019}
Stephen Robertson and Hugo Zaragoza. 2009.
\newblock \href {https://doi.org/10.1561/1500000019} {The probabilistic relevance framework: Bm25 and beyond}.
\newblock \emph{Found. Trends Inf. Retr.}, 3(4):333–389.

\bibitem[{Rush et~al.(2015)Rush, Chopra, and Weston}]{Rush_2015}
Alexander~M. Rush, Sumit Chopra, and Jason Weston. 2015.
\newblock \href {https://doi.org/10.18653/v1/d15-1044} {A neural attention model for abstractive sentence summarization}.
\newblock \emph{Proceedings of the 2015 Conference on Empirical Methods in Natural Language Processing}.

\bibitem[{Venkit et~al.(2024)Venkit, Laban, Zhou, Mao, and Wu}]{venkit2024searchenginesaiera}
Pranav~Narayanan Venkit, Philippe Laban, Yilun Zhou, Yixin Mao, and Chien-Sheng Wu. 2024.
\newblock \href {https://arxiv.org/abs/2410.22349} {Search engines in an ai era: The false promise of factual and verifiable source-cited responses}.
\newblock \emph{Preprint}, arXiv:2410.22349.

\bibitem[{Xiang et~al.(2024)Xiang, Wu, Zhong, Wagner, Chen, and Mittal}]{xiang2024certifiablyrobustragretrieval}
Chong Xiang, Tong Wu, Zexuan Zhong, David Wagner, Danqi Chen, and Prateek Mittal. 2024.
\newblock \href {https://arxiv.org/abs/2405.15556} {Certifiably robust rag against retrieval corruption}.
\newblock \emph{Preprint}, arXiv:2405.15556.

\bibitem[{Xiao et~al.(2023)Xiao, Liu, Zhang, and Muennighoff}]{bge_embedding}
Shitao Xiao, Zheng Liu, Peitian Zhang, and Niklas Muennighoff. 2023.
\newblock \href {https://arxiv.org/abs/2309.07597} {C-pack: Packaged resources to advance general chinese embedding}.
\newblock \emph{Preprint}, arXiv:2309.07597.

\bibitem[{Yalçın and Köse(2024)}]{seoaeo}
Nursel Yalçın and Utku Köse. 2024.
\newblock \href {https://www.forbes.com/councils/forbesbusinesscouncil/2023/03/14/the-future-of-seo-is-answer-engine-optimization-aeo/} {The future of seo is answer engine optimization (aeo)}.
\newblock \emph{Forbes}.

\bibitem[{Yang et~al.(2018)Yang, Qi, Zhang, Bengio, Cohen, Salakhutdinov, and Manning}]{yang2018hotpotqa}
Zhilin Yang, Peng Qi, Saizheng Zhang, Yoshua Bengio, William~W. Cohen, Ruslan Salakhutdinov, and Christopher~D. Manning. 2018.
\newblock {HotpotQA}: A dataset for diverse, explainable multi-hop question answering.
\newblock In \emph{Conference on Empirical Methods in Natural Language Processing ({EMNLP})}.

\bibitem[{Yates et~al.(2021)Yates, Nogueira, and Lin}]{yates-etal-2021-pretrained}
Andrew Yates, Rodrigo Nogueira, and Jimmy Lin. 2021.
\newblock \href {https://doi.org/10.18653/v1/2021.naacl-tutorials.1} {Pretrained transformers for text ranking: {BERT} and beyond}.
\newblock In \emph{Proceedings of the 2021 Conference of the North American Chapter of the Association for Computational Linguistics: Human Language Technologies: Tutorials}, pages 1--4, Online. Association for Computational Linguistics.

\bibitem[{Zaoad et~al.(2025)Zaoad, Zawad, Ranade, Krogman, Khan, and Holt}]{zaoad2025graphbasedrerankingemergingtechniques}
Md~Shahir Zaoad, Niamat Zawad, Priyanka Ranade, Richard Krogman, Latifur Khan, and James Holt. 2025.
\newblock \href {https://arxiv.org/abs/2503.14802} {Graph-based re-ranking: Emerging techniques, limitations, and opportunities}.
\newblock \emph{Preprint}, arXiv:2503.14802.

\bibitem[{Zhang et~al.(2025)Zhang, Xin, Fang, Liu, Yi, Li, and Liu}]{10.1145/3696410.3714756}
Baolei Zhang, Haoran Xin, Minghong Fang, Zhuqing Liu, Biao Yi, Tong Li, and Zheli Liu. 2025.
\newblock \href {https://doi.org/10.1145/3696410.3714756} {Traceback of poisoning attacks to retrieval-augmented generation}.
\newblock In \emph{Proceedings of the ACM on Web Conference 2025}, WWW '25, page 2085–2097, New York, NY, USA. Association for Computing Machinery.

\bibitem[{Zhang et~al.(2022)Zhang, Long, Xu, and Xie}]{Zhang2022HLATREM}
Yanzhao Zhang, Dingkun Long, Guangwei Xu, and Pengjun Xie. 2022.
\newblock Hlatr: Enhance multi-stage text retrieval with hybrid list aware transformer reranking.
\newblock \emph{ArXiv}, abs/2205.10569.

\bibitem[{Zhao et~al.(2024)Zhao, Khazanchi, Xing, He, Xu, and Lane}]{zhao2024-api-attacks}
Wanru Zhao, Vidit Khazanchi, Haodi Xing, Xuanli He, Qiongkai Xu, and Nicholas~Donald Lane. 2024.
\newblock Attacks on third-party apis of large language models.
\newblock In \emph{ICLR 2024 Workshop on Secure and Trustworthy Large Language Models}.

\bibitem[{Zhong et~al.(2023)Zhong, Huang, Wettig, and Chen}]{zhong-etal-2023-poisoning}
Zexuan Zhong, Ziqing Huang, Alexander Wettig, and Danqi Chen. 2023.
\newblock \href {https://doi.org/10.18653/v1/2023.emnlp-main.849} {Poisoning retrieval corpora by injecting adversarial passages}.
\newblock In \emph{Proceedings of the 2023 Conference on Empirical Methods in Natural Language Processing}, pages 13764--13775, Singapore. Association for Computational Linguistics.

\bibitem[{Zhou et~al.(2025)Zhou, Lee, Zhan, Chen, Li, Wang, Haddadi, and Yilmaz}]{zhou2025trustragenhancingrobustnesstrustworthiness}
Huichi Zhou, Kin-Hei Lee, Zhonghao Zhan, Yue Chen, Zhenhao Li, Zhaoyang Wang, Hamed Haddadi, and Emine Yilmaz. 2025.
\newblock \href {https://arxiv.org/abs/2501.00879} {Trustrag: Enhancing robustness and trustworthiness in retrieval-augmented generation}.
\newblock \emph{Preprint}, arXiv:2501.00879.

\bibitem[{Zou et~al.(2024)Zou, Geng, Wang, and Jia}]{zou2024poisonedrag}
Wei Zou, Runpeng Geng, Binghui Wang, and Jinyuan Jia. 2024.
\newblock \href {https://arxiv.org/abs/2402.07867} {Poisonedrag: Knowledge poisoning attacks to retrieval-augmented generation of large language models}.
\newblock \emph{Preprint}, arXiv:2402.07867.

\end{thebibliography}

\appendix
\clearpage

\section{Prompts to LLM}
\label{sec:appendix1}
\cref{fig:Prompt_example} shows the two prompts we use to query the LLMs.
\begin{figure}
\begin{tcolorbox}[title=Prompts given to LLMs]
\textbf{Prompt 1: }You are a helpful assistant, below is a query from a user and some relevant contexts. 
Answer the question given the information in those contexts / keywords. Your answer should be short ,concise and without explanations ,just answer with one or two words like yes or no. 
If you cannot find the answer to the question, just say "I don't know". 
Contexts / Keywords: [context / keywords] Query: [question] Answer:
\\
\textbf{Prompt 2: }You are a helpful assistant, below is a query from a user and some relevant contexts. 
Answer the question given the information in those contexts.
If you cannot find the answer to the question, just say "I don't know". 
Contexts: [context] Query: [question] Answer:

\textbf{Prompt 3: }You are a helpful assistant, Answer the query with no more than ten words. If you do not know the answer confidently, just say "I don't know". Query: [question] Answer: 

\end{tcolorbox} 
\caption{Example of prompts given to LLM, Prompt 1 is used in all of the defense methods to generate the final output. Prompt 2 is only used in the phase to generate keywords and Prompt 3 is used for No-RAG from the paper~\cite{xiang2024certifiablyrobustragretrieval}.} 

\label{fig:Prompt_example}
\end{figure}

\section{Deconstructing PoisonedRAG}
\label{sec:deconstry}
\begin{figure*}[htbp]
\begin{tcolorbox}[title=PoisonedRAG Example]
\textbf{Question:}  "When did the apple iphone se come out?"
\\
\textbf{Benign documents 1: }"March 21, 2017 iPhone SE shipped with iOS 9.3, supporting iPhone 6 and 6S-exclusive features such as Apple Pay, Live Photos, Retina Flash and always-on Siri voice activation.[6][10][16]"
\\
\textbf{Benign documents 2: }"On launch, it was released in models with either 16 or 64 GB of internal storage; on March 21, 2017, Apple announced that these models would be replaced by new improved models at the same launch price points, releasing on March 24, 2017.[11][12]"
\\
\color{red}{\textbf{Poisoned Document 1: }"When did the apple iphone se come out. The Apple archive, housing important launch events, mentions a significant one on April 15, 2015. On this day, the compact, budget-friendly iPhone SE hit the market, impressing many gadget enthusiasts."}
\\
\color{red}{\textbf{Poisoned Document 2: }"When did the apple iphone se come out. The iPhone series has released several models. Notable amongst them is the iPhone SE, introduced to the market on April 15, 2015, attracting considerable consumer attention."}

\end{tcolorbox} 
\caption{Example of PoisonedRAG attacks. Poisoned documents injected into the database are all very similar to each other and focus solely on ensuring similarity to the query, the similarities among the retrieved documents are never considered.} 

\label{fig:PoisonedRAG_example}
\end{figure*}
PoisonedRAG~\cite{zou2024poisonedrag} is an adversarial attack on RAG systems that operates in two stages.
The first part enhances the semantic similarity of the adversarial passage to the query, increasing the likelihood of it being retrieved.
The second part introduces adversarial content to mislead the model into generating a specific incorrect response.
%

While the approach used to achieve the first part of the attack is effective, it is also simple.
Specifically, the adversarial passage is constructed by prepending the query into the poisonous passage.
Despite its simplicity, PoisonedRAG degrades the accuracy of the LLMs significantly.
As shown in \cref{tab:pos_attack} (first row), the attack achieves an ASR of 54.3\% on average across three datasets with just one adversarial passage retrieved as the most similar to the query.

\begin{table}[t]
\resizebox{\columnwidth}{!}{%
\begin{tabular}{lcccc}
\toprule
\textbf{Attack Method}                                 & \textbf{HotpotQA} & \textbf{NQ} & \textbf{MS-MARCO} & \textbf{Average} \\
\midrule
Normal retrieved                          & 59.0              & 56.0        & 48.0              & 54.3              \\
\quad w/o question & 66.0              & 61.0        & 51.0              & 59.3              \\
\midrule
Poisoned in the middle                    & 59.0              & 54.0        & 37.0              & 50.0              \\
\quad w/o question & 63.0              & 51.0        & 34.0              & 49.3              \\
\bottomrule
\end{tabular}%
}
\caption{PoisonedRAG Attack Success Rate (\%) where the retrieval part is removed, and the poisoned documents are placed in the middle.}
\label{tab:pos_attack}
\end{table}

Our analysis reveals that the prepended query in the adversarial passage does not significantly affect the ASR.
As shown in \cref{tab:pos_attack} (second row), removing the prepended query leads to an increase in the ASR.
This shows that the query was prepended only to ensure that the retriever retrieves the adversarial document, but not affecting the accuracy significantly.
Furthermore, \cref{tab:pos_attack} (third and fourth row) shows that the position of the poisoned document within the retrieved documents set influences the ASR significantly, with a decrease in average ASR of 10\%.
This phenomenon is similar to the lost-in-the-middle effect~\cite{liu-etal-2024-lost}, where the position of the document impacts its effectiveness in influencing the output of the reader model.

Due to its straightforward approach of prepending the query to the adversarial documents, PoisonedRAG attacks can be easily identified.
As demonstrated in \cref{fig:Bm25_heatmap} and \cref{fig:PoisonedRAG_example}, the attacks injected into the database often exhibit considerable similarity to one another.
By focusing on the similarities between the documents in the retrieved set, we can filter out adversarial passages and decrease the ASR.

\begin{figure*}[t]
    \centering
    \begin{subfigure}{0.48\textwidth}
        \includegraphics[width=\linewidth]{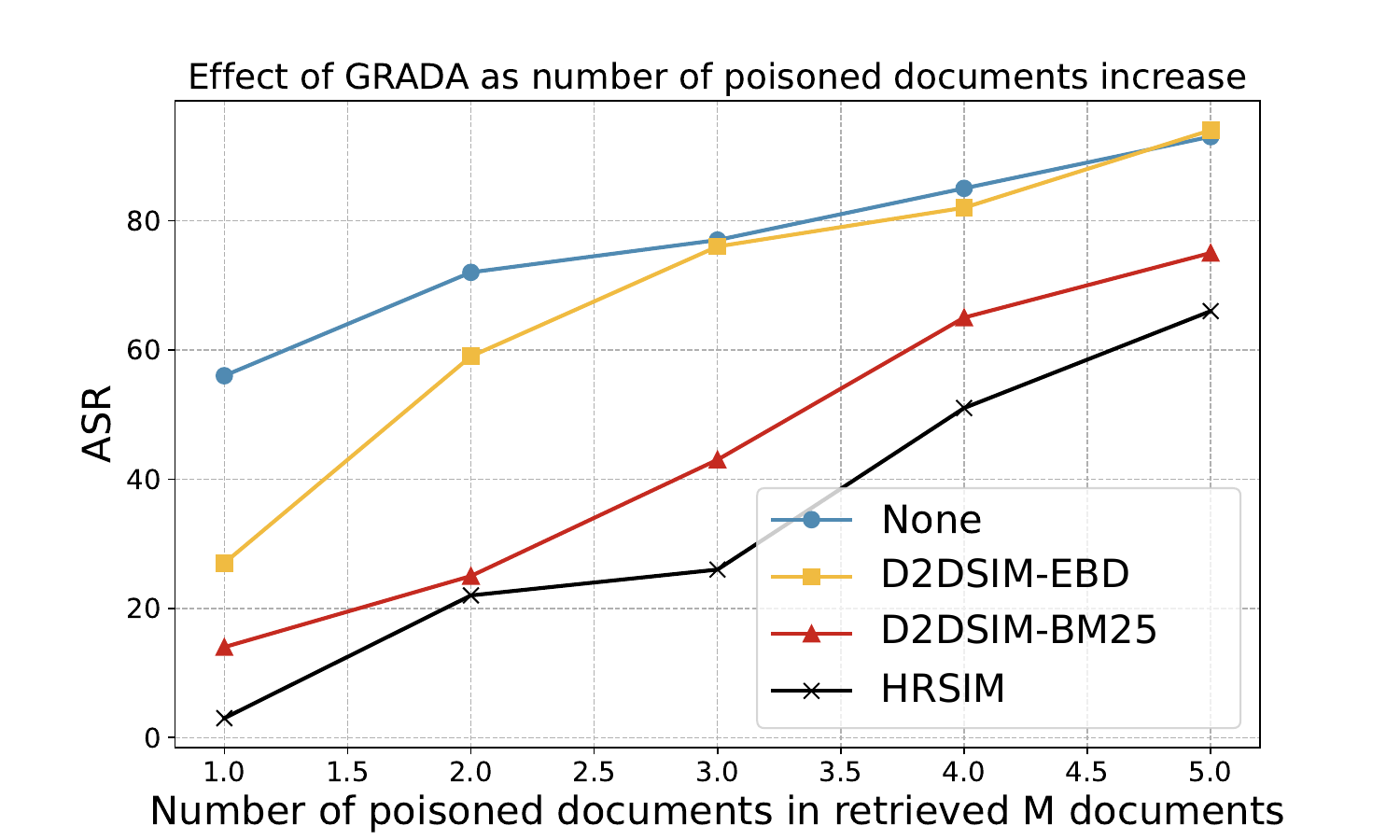}
        \caption{ASR of GRADA as poisoned documents increase.}
        \label{fig:Effect_GRADA}
    \end{subfigure}
    \hfill
    \begin{subfigure}{0.48\textwidth}
        \includegraphics[width=\linewidth]{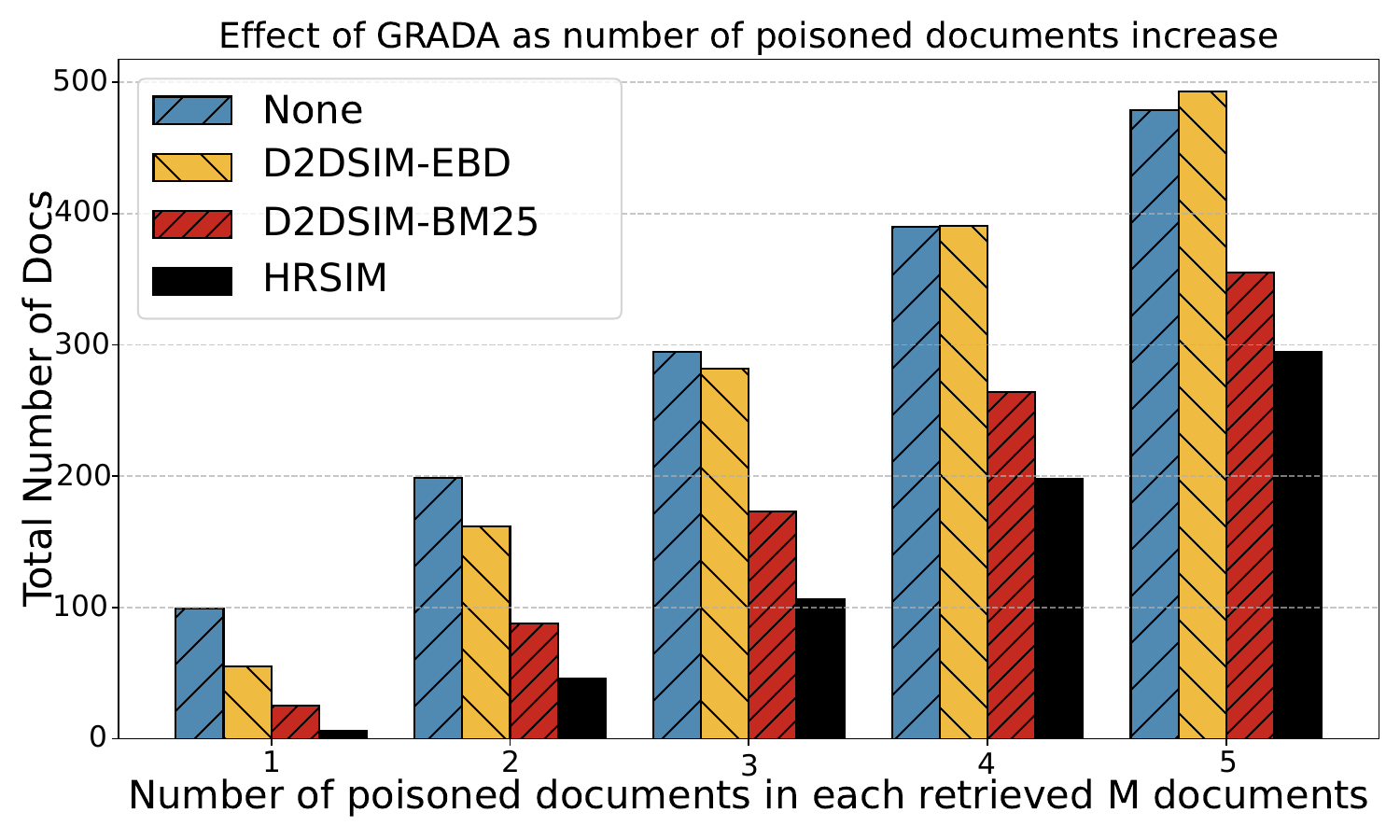}
        \caption{Total poisoned documents remain after filtering.}
        \label{fig:Effect_GRADA_num}
    \end{subfigure}
    \caption{Impact of increasing poisoned documents on GRADA's performance in NQ dataset (GPT-3.5-Turbo, $M=10$).}
    \label{fig:Effect_GRADA_combined}
\end{figure*}

\section{Ablation Study}
\subsection{Number of poisoned documents increase}
\label{app:pos_increase}

As shown in \cref{fig:Effect_GRADA}, the effectiveness of \ourmethod reduces as the proportion of poisoned documents increases.
When using D2DSIM-EBD, the ASR achieved by \ourmethod approaches that of an undefended system.
However, HRSIM remains effective, achieving a 27\% reduction in ASR even when half of the retrieved documents are adversarial.
This is further supported by \cref{fig:Effect_GRADA_num}, which shows that 38\% of poisoned documents are still successfully filtered.

\begin{figure}[t]
    \centering
    \includegraphics[width=\linewidth]{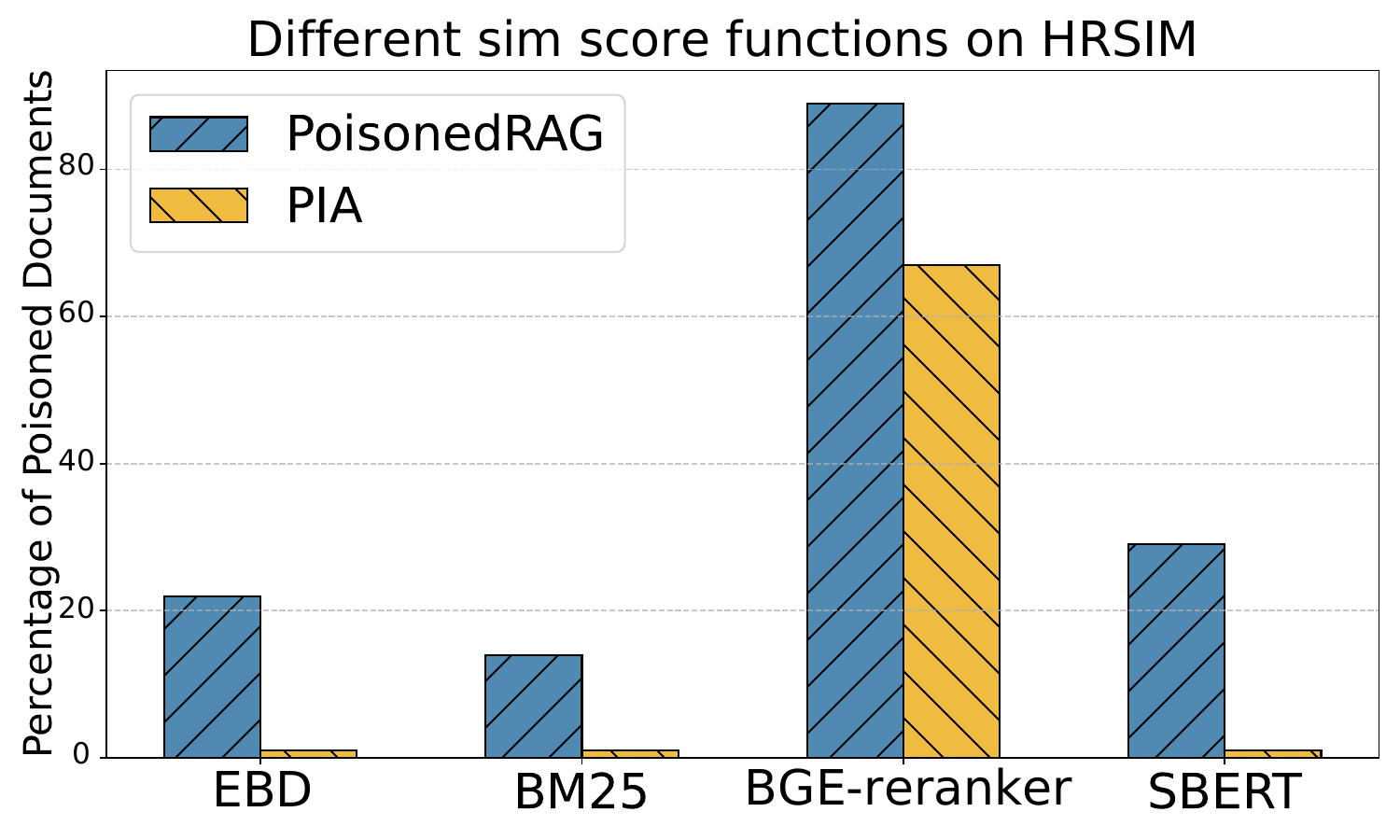}
    \caption{HRSIM performance with different similarity functions selection on MSMARCO dataset. The figure illustrates the proportion of test instances in which poisoned documents remain among the top five retrieved results.}
    \label{fig:Diff_sim}
    \vspace{-4mm}
\end{figure}
\begin{figure}[t]
    \centering
    \includegraphics[width=\linewidth]{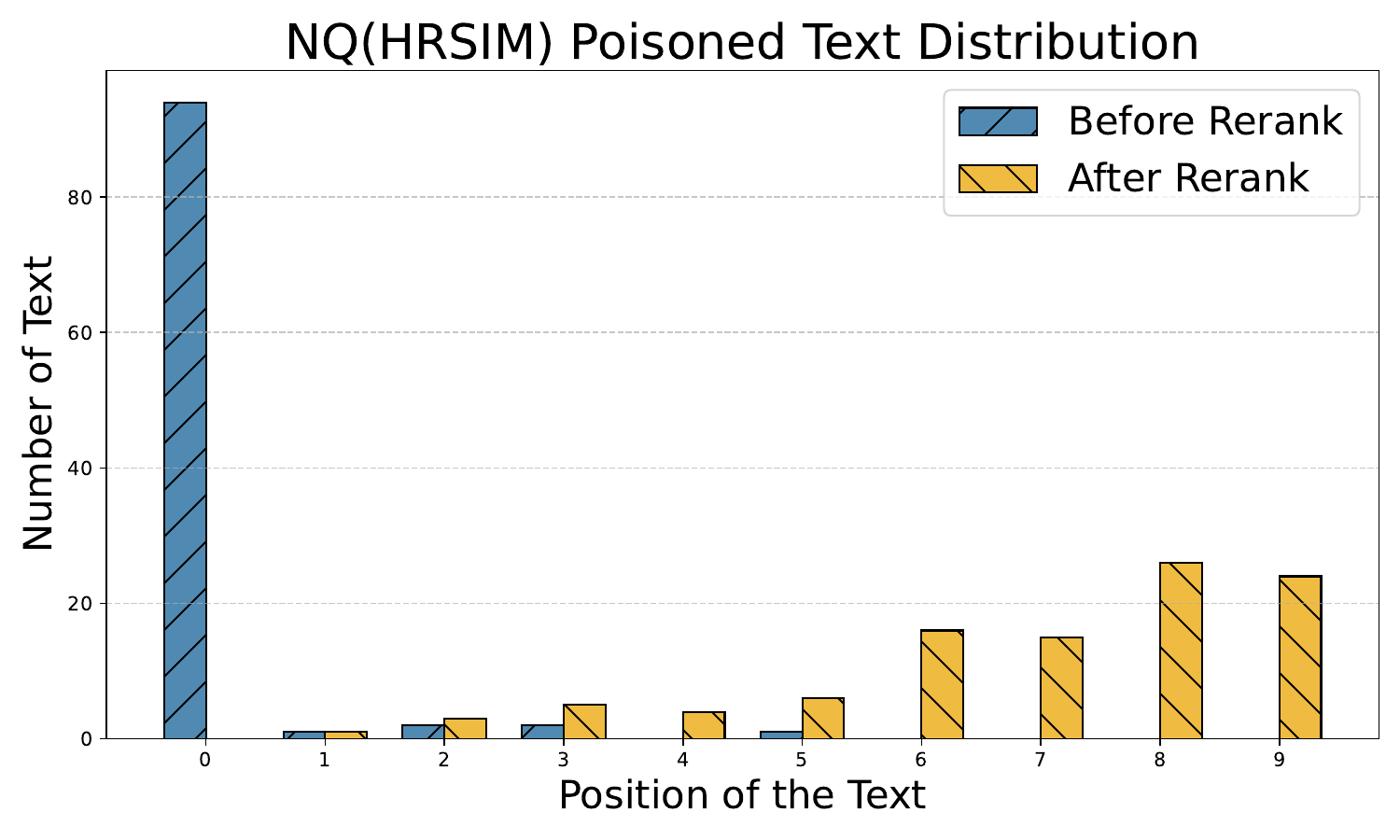}
    \caption{Distribution of poisoned document positions after applying \ourmethod(HRSIM) in the NQ dataset. Documents positioned below rank 5 are effectively mitigated by the ranking algorithm. Other results are showed in \cref{fig:ground_NQ_combined,tab:NQ_rerank,tab:MS_rerank,tab:Hotpot_rerank}}
    \label{fig:position_MSMARCO(BM25)}
\end{figure}

\subsection{Selections of HRSIM.}

\label{sec:side_studies}

Thus far, our focus has primarily been on utilizing BM25 for HRSIM. In this section, we explore other similarity functions for HRSIM. As shown in \cref{fig:Diff_sim}, we extend our analysis by incorporating SBERT~\cite{reimers-gurevych-2019-sentence}, alongside the three previously discussed methods, to better capture document-to-document similarity. Our results indicate that both EBD and SBERT exhibit strong overall performance against PIA and PoisonedRAG attacks. In contrast, BGE-Reranker struggles to effectively filter out poisoned documents, likely due to its primary training objective of computing query-to-document similarities rather than document-to-document relationships. HRSIM, when combined with BM25, effectively minimize the presence of poisoned documents, reducing them to just 14 out of 100 test instances. This outcome underscores its remarkable effectiveness in filtering malicious content.

\subsection{Different initial score vector}
\begin{figure*}[t]
    \centering
    \begin{subfigure}{0.48\textwidth}
        \includegraphics[width=\linewidth]{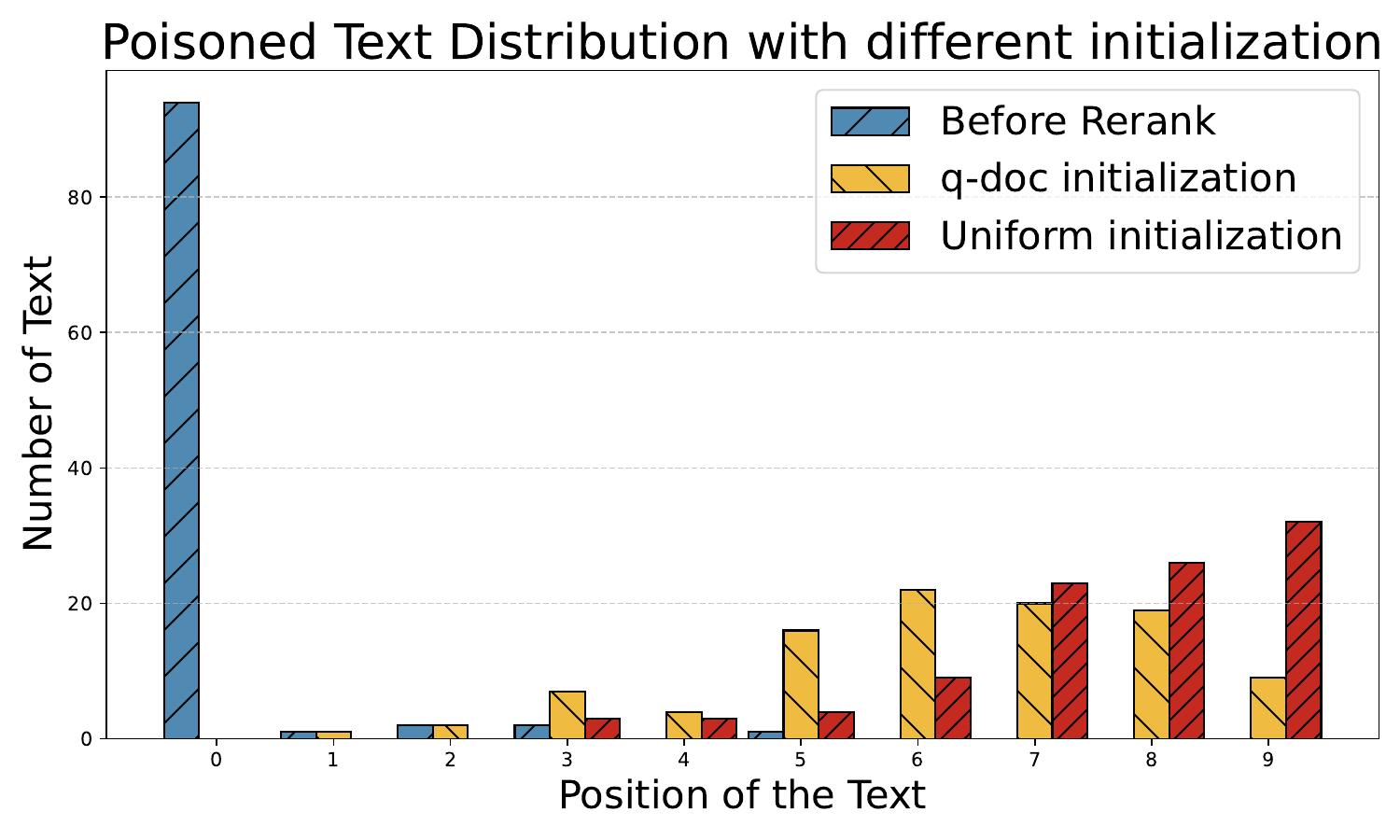}
        \caption{Distribution of Poisoned document positions after applying \ourmethod (HRSIM) with different initialization in the NQ dataset.}
        \label{fig:NQ(PS)_INIT}
    \end{subfigure}
    \hfill
    \begin{subfigure}{0.48\textwidth}
        \includegraphics[width=\linewidth]{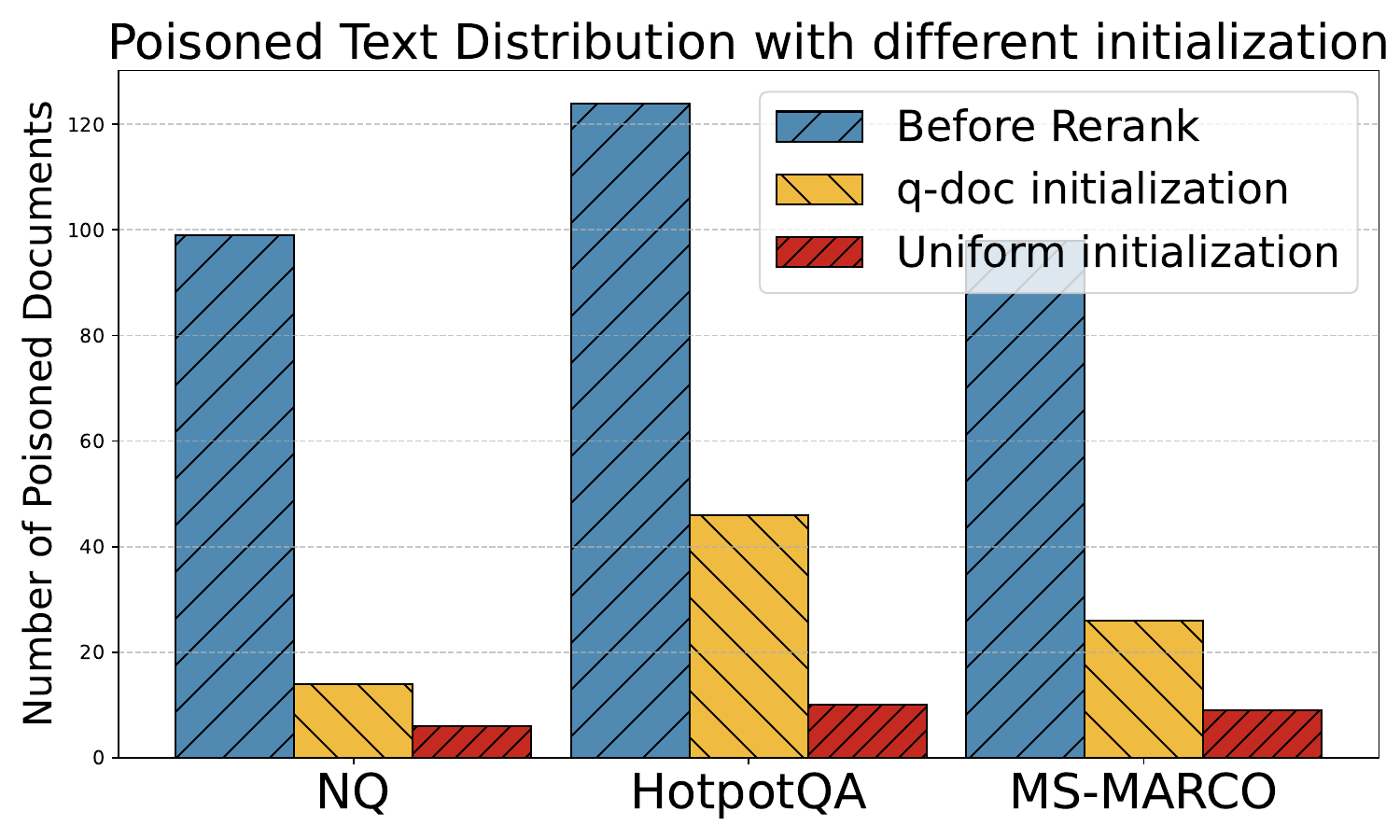}
        \caption{Total number of poisoned documents after applying \ourmethod (HRSIM) with different initialization in the NQ dataset.}
        \label{fig:All_three_INIT}
    \end{subfigure}
    \caption{Impact of different initialization score vectors on GRADA's performance ($M=10$).}
    \label{fig:INIT_combined}
\end{figure*}

Different initial score vectors can have a significant impact on the final distribution of documents in certain cases.
For instance, we experimented with initializing the score vector with query-document similarity $s^* = \left[ \frac{sim(q,v_0)}{\sum_{j=0}^{n} sim(q,v_j)},\frac{sim(q,v_1)}{\sum_{j=0}^{n} sim(q,v_j)},...,\frac{sim(q,v_n)}{\sum_{j=0}^{n} sim(q,v_j)} \right]$.
As shown in \cref{fig:NQ(PS)_INIT}, using a query-document initialization results in more documents being positioned between rank 5 and 8, rather than lower.
We hypothesize that this is because adversarial documents may receive disproportionately high initial scores compared to benign documents.
Such an imbalance gives adversarial documents a substantial advantage, particularly when the edge weights between documents are relatively small.
In these scenarios, the graph-based reranking process may struggle to compensate for this initial disparity, as illustrated in \cref{fig:initial_draw}.
From the analysis in \cref{fig:All_three_INIT}, we observe that this phenomenon is more prevalent in datasets like HotpotQA.
\label{app:initial_score_vector}

\subsection{Ranking distribution.}

We have demonstrated the effectiveness of our approach in enhancing defense performance. To gain a deeper understanding of its impact, we further analyze how our method systematically lowers the ranking of poisoned documents. As illustrated in \cref{fig:position_MSMARCO(BM25)}, the position distribution of poisoned documents within the retrieval set shifts significantly after applying \ourmethod with D2DSIM-BM25. Notably, over 70\% of poisoned documents are relegated beyond the top five positions, substantially reducing their influence. These findings confirm that \ourmethod is both robust and effective in mitigating adversarial attacks.

\begin{table*}[t]
\centering
\resizebox{0.95\linewidth}{!}{%
\begin{tabular}{@{}lcccc@{}}
\toprule
\textbf{Defense} & \textbf{Total Time (s)} & \textbf{Processing Time (s)} & \textbf{Defense-Only Time (s)} & \textbf{Defense-Only Processing (s)} \\
\midrule
Keyword Aggregation  & 12.61 & 11.11 & 11.59 & 9.21\\
\ourmethod (D2DSIM-EBD)            & 1.56 & 1.12 & 0.62 & 0.62\\
\ourmethod (D2DSIM-BM25)              & 0.97 & 0.52 & 0.02 & 0.02\\
\ourmethod (HRSIM) & 1.05 & 0.61 & 0.05 & 0.05 \\
                               \bottomrule

\end{tabular}
}
\caption{ Runtime Comparison (on GPT-3.5-Turbo, average per query): Total Time (s) and Processing Time (s) represent the complete runtime for answering one question, including retrieval, defense method, and LLM response generation. In contrast, Defense-Only Time (s) measures exclusively the runtime of the defense methods themselves. Total Time is recorded using Python's time.time() function, whereas Processing Time is measured with Python's time.process\_time() function.}
\label{tab:time}
\end{table*}

\subsection{Different Retriever}
To further assess the generality of our approach, we conduct experiments on bge-small-en-v1.5~\cite{bge_embedding}. As shown in \cref{tab:extra_r}, the overall trends remain consistent with our main experiments. In the no-attack setting, all defenses incur only a small EM drop compared to the baseline retriever, suggesting limited utility loss. Under PoisonedRAG, the baseline suffers an ASR of 41.3\% with a large EM drop, while our defenses substantially reduce ASR and recover EM—HRSIM achieves the best trade-off with an ASR of only 5.3\% and the highest EM of 62.3\%. For PIA, the baseline reaches 31.0\% ASR, but our methods again nearly eliminate the attack, with D2DSIM-EBD completely blocking it while retaining 59.0\% EM. These results demonstrate that our defenses transfer effectively across retrievers and maintain robustness against different poisoning strategies.

\begin{table*}[t]
\centering
\resizebox{0.9\linewidth}{!}{%
\begin{tabular}{@{}lccc@{}}
\toprule
\multirow{2}{*}{\textbf{Defense}} &
\textbf{No Attack} &
\textbf{PoisonedRAG} &
\textbf{PIA} \\
& \small{ASR $\downarrow$ / EM $\uparrow$} 
& \small{ASR $\downarrow$ / EM $\uparrow$} 
& \small{ASR $\downarrow$ / EM $\uparrow$} \\
\midrule
None                     & -- / 65.3$\pm$0.5  & 41.3$\pm$1.2 / 44.0$\pm$0.8 & 31.0$\pm$0.0 / 43.7$\pm$0.5 \\
\ourmethod (D2DSIM-EBD)  & -- / 59.7$\pm$0.5  & 28.3$\pm$0.5 / 46.7$\pm$0.5 & 0.0$\pm$0.0 / 59.0$\pm$0.0 \\
\ourmethod (D2DSIM-BM25) & -- / 60.7$\pm$0.9  & 17.7$\pm$0.5 / 56.3$\pm$0.5 & 8.0$\pm$0.0 / 55.7$\pm$0.5 \\
\ourmethod (HRSIM)       & -- / 59.3$\pm$0.9  & 5.3$\pm$0.5  / 62.3$\pm$1.2 & 1.0$\pm$0.0 / 58.3$\pm$0.5 \\
\bottomrule
\end{tabular}
}
\caption{Performance of defenses on an bge-small-en-v1.5 under \textbf{No Attack}, \textbf{PoisonedRAG}, and \textbf{PIA}.}
\label{tab:extra_r}
\vspace{-4mm}
\end{table*}

\subsection{Computational Complexities.}
The overall complexity of \ourmethod consists of two main components:
\begin{itemize}
    \item \textbf{Similarity matrix construction:} $O(N^2)$, where N is the number of retrieved documents. This step can incur additional costs depending on the chosen similarity function. For example, using D2DSIM-EBD (embedding-based document similarity), the complexity becomes $O(N^2 \cdot d)$, where $d$ is the embedding dimension. BM25-based similarity: the complexity is $O(N^2 \cdot L)$, where L is the average document length. This is efficient due to the sparsity of token overlaps and inverted index optimizations. Here, since we are reranking the documents after the retrieval step. The retrieved documents set is usually constrained with limited amounts of data, making this a viable solution.
    \item \textbf{Graph-based reranking (e.g., PageRank):} $O(n+m)$, where n is the number of nodes (documents) and m is the number of edges in the constructed similarity graph.

\end{itemize}

The only defense Keyword Aggregation requires querying the language model N times—once per document—to collect individual answers before aggregating: $O(N*C_{LM})$ (where $C_{LM}$ refers to the language model’s cost). This incurs significantly higher costs in terms of API calls and model generation time, especially with large models.

GRADA, by comparison, does not require any model calls. The only required model call is after GRADA to query the final answer, making it more efficient and scalable for large-scale or production RAG deployments. 

\section{Different initial score vector demonstration}
\begin{figure*}[htbp]
\begin{tcolorbox}[title=Initial Score Example in \cref{fig:initial_draw}]
\textbf{Question:}  "Are Random House Tower and 888 7th Avenue both used for real estate?"
\\
\textbf{Documents 1: }"The former Bertelsmann Building, now known as 1540 Broadway, is a 44-story, 733 foot (223 m) office tower in Times Square in Manhattan..."
\\
\textbf{Documents 2: }"The Random House Tower, also known as the Park Imperial Apartments, is a 52-story mixed-use tower in New York City, United States, that is..."
\\
\textbf{Documents 3: }"888 7th Avenue is a 628 ft (191m) tall modern-style office skyscraper in Midtown Manhattan which was completed in 1969 and has 46 floors. Emery Roth \& Sons designed..."
\\
\textbf{Documents 4: }"What do the estates of film stars Vincent Price and Glenn Ford have in common? And what do each of these estates have in common with valuables owned by Laugh-In’s Arte..."
\\
\textbf{Documents 5: }"750 Seventh Avenue is a 615 ft (187m) tall Class-A office skyscraper in New York City. It was completed in 1989 in the postmodern style and has 36 floors..."
\\
\textbf{Documents 6: }"The Fisk Towers is a front for the Kingpin (Wilson Fisk)'s public ventures as well as a base of operations for his criminal activities, until..."
\\
\color{red}{\textbf{Document 0: }"Are Random House Tower and 888 7th Avenue both used for real estate?.Random House Tower is occupied by a publishing company, not devoted to real estate. 888 7th Avenue is primarily used for law firms, again not real estate operations."}

\end{tcolorbox} 
\caption{Document examples used to generate \cref{fig:initial_draw} to demonstrate different initial score vector and their results when the adversarial documents receive
significantly higher initial scores compared to benign documents. Red Documents indicates the poisoned document.} 

\label{fig:ranking_example}
\end{figure*}
\begin{figure*}
    \centering
    \includegraphics[width=\linewidth]{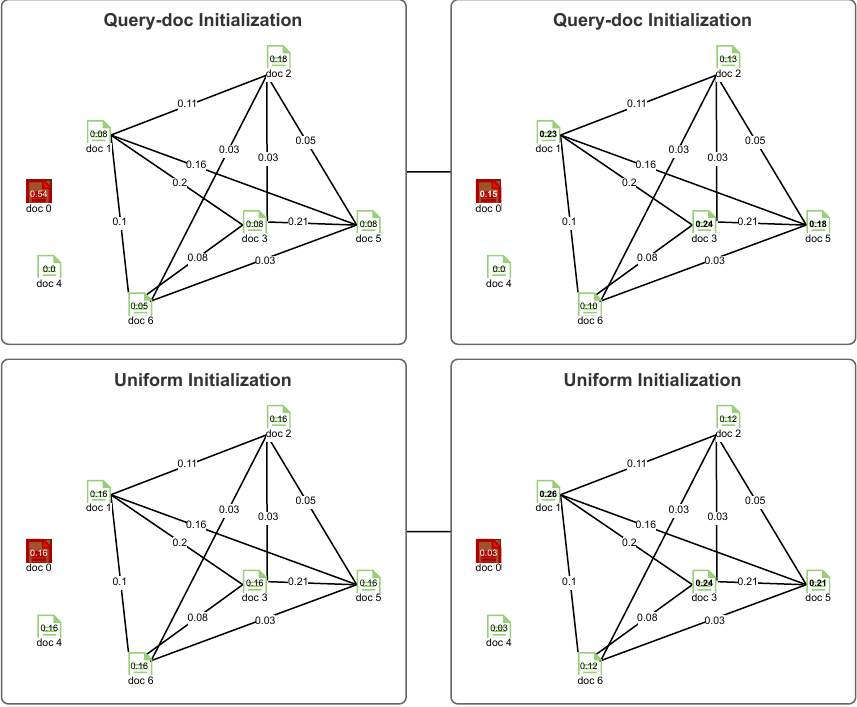}
    \caption{A demonstration on different initial score vector and their results when the adversarial documents receive significantly higher initial scores compared to benign documents. This is an example from the HotpotQA dataset with the question:" Are Random House Tower and 888 7th Avenue both used for real estate?". The top 4 ranked documents are listed with bold final values.}
    \label{fig:initial_draw}
\end{figure*}

\cref{fig:ranking_example} shows the documents we used in \cref{fig:initial_draw}.

\section{Computational Resources}

The cost of a single defense run on GPT-3.5-Turbo is \$0.50, identical to a standard query since the method does not introduce additional API calls.
Experiments for LLaMA-3 and Qwen2.5 were conducted on a single NVIDIA A100 80GB GPU, with each defense run taking one hour to complete.

\section{License and Distribution Terms}
\label{app:license}
The dataset used in our experiments is publicly available under Creative Commons Attribution 4.0 International (MS-MARCO) and Apache License 2.0 (NQ, HotpotQA).
The code used in our experiments is publicly available under MIT License (BM25s, PoisonedRAG)~\url{https://anonymous.4open.science/r/GRADA-266D}

\begin{figure*}[t]
    \centering
    \begin{subfigure}{0.32\textwidth}
        \includegraphics[width=\linewidth ]{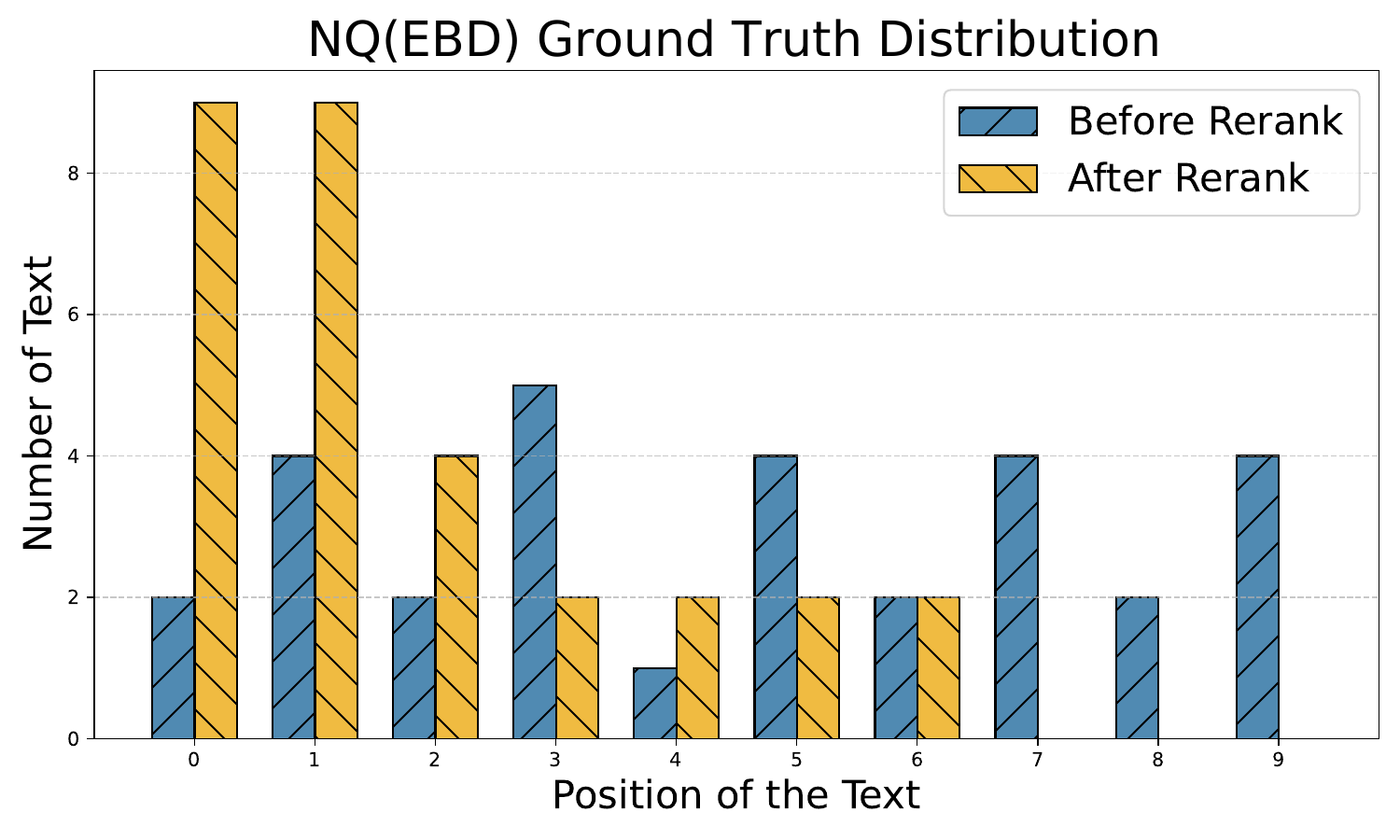}
        \caption{D2DSIM-EBD}
        \label{fig:ground_NQ_cos}
    \end{subfigure}
    \begin{subfigure}{0.32\textwidth}
        \includegraphics[width=\linewidth]{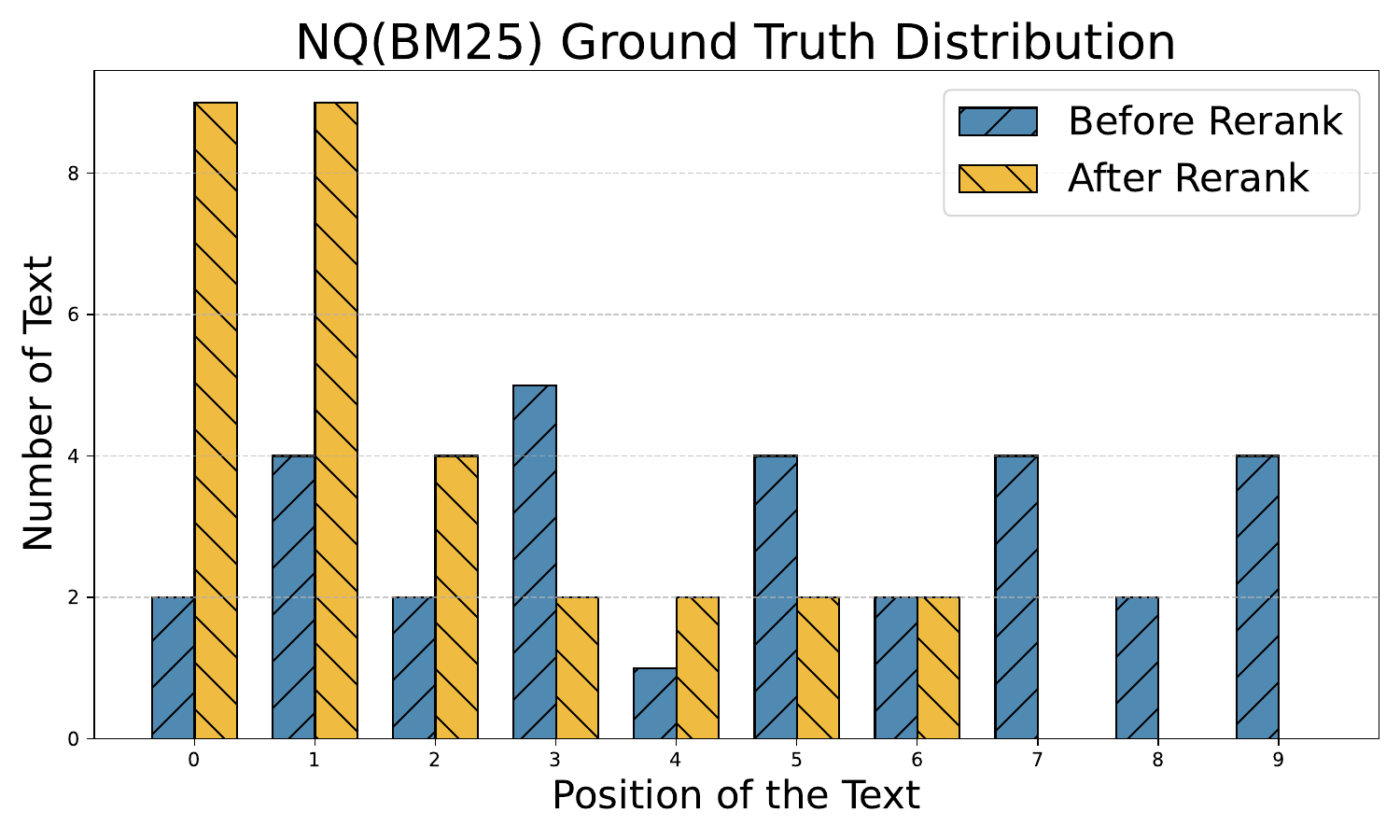}
        \caption{D2DSIM-BM25}
        \label{fig:ground_NQ_bm25}
    \end{subfigure}
    \begin{subfigure}{0.32\textwidth}
        \includegraphics[width=\linewidth]{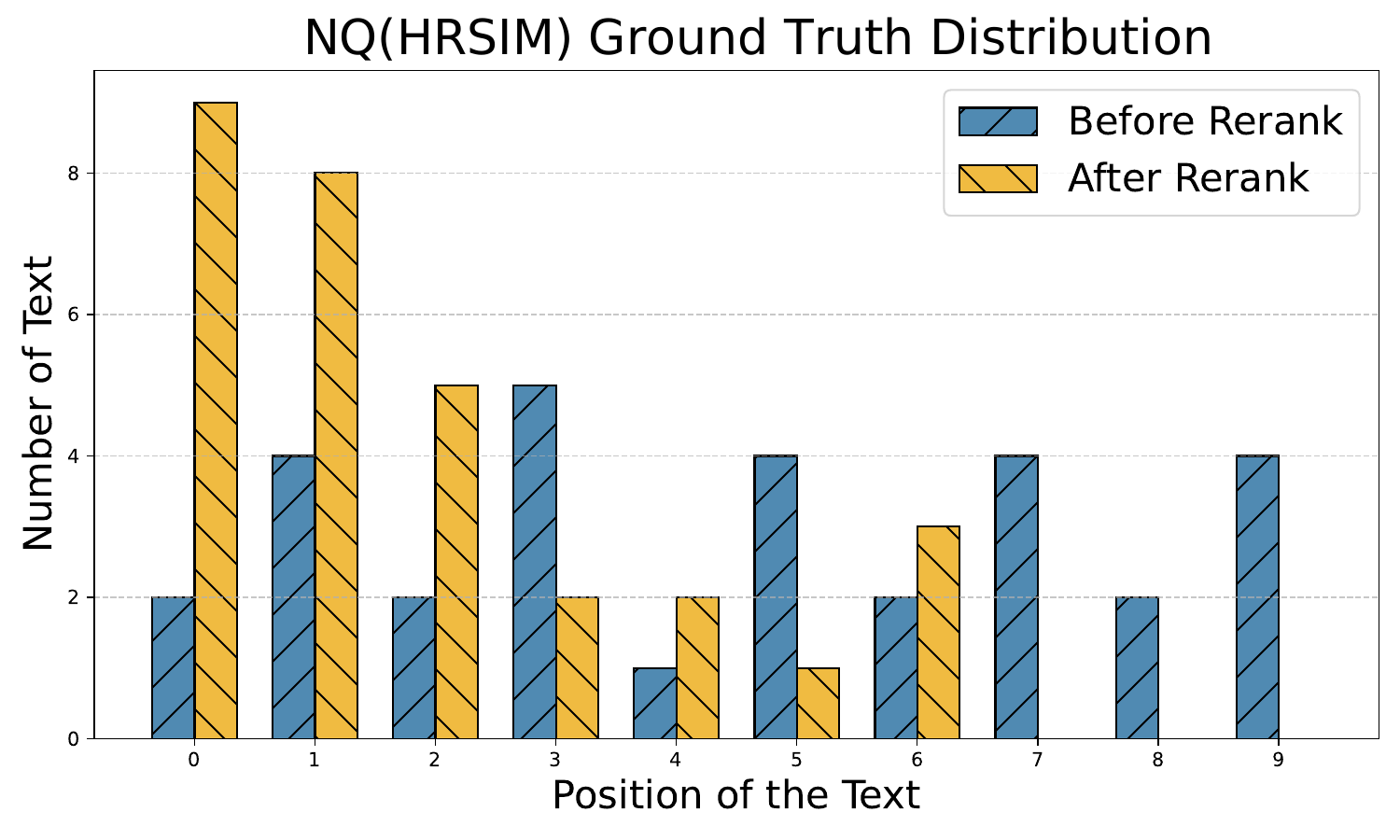}
        \caption{HRSIM}
        \label{fig:ground_NQ_ps}
    \end{subfigure}
    \caption{Distribution of Ground Truth document positions after applying \ourmethod in the NQ dataset with different ranking methods.}
    \label{fig:ground_NQ_combined}
\end{figure*}

\begin{table*}
\resizebox{\linewidth}{!}{%
\begin{tabular}{lcccc}
\toprule
                          \textbf{Defense Method} & \textbf{PoisonedRAG} & \textbf{PoisonedRAG(Hotflip)} & \textbf{PIA} & \textbf{Phantom} \\
\midrule
No Defense                 & 99.0  & 99.0  & 96.0 & 94.0  \\
\midrule
HLATR                      & 100.0 & 100.0 & 93.0 & 89.0  \\
BGE-reranker               & 100.0 & 98.0  & 47.0 & 58.0  \\
\ourmethod (D2DSIM-EBD)    & 55.0  & 20.0  & 6.0  & 5.0   \\
\ourmethod (D2DSIM-BM25)   & 25.0  & 16.0  & 6.0  & 4.0   \\
\ourmethod (HRSIM)         & 13.0  & 8.0   & 7.0  & 2.0   \\
\bottomrule
\end{tabular}%
}
\caption{The percentage of poisoned documents in the given context to LLM before and after different defense methods on the NQ dataset. A lower value is better. Method Keyword not included as it does not conduct reranking.}
\label{tab:NQ_rerank}
\end{table*}

\begin{table*}[]
\resizebox{\linewidth}{!}{%
\begin{tabular}{lcccccccc}
\toprule
                          \textbf{Defense Method} & \textbf{PoisonedRAG} & \textbf{PoisonedRAG(Hotflip)} & \textbf{PIA} & \textbf{Phantom} \\
\midrule
No Defense               & 98.0  & 99.0  & 89.0  & 65.0 \\
\midrule
HLATR                    & 98.0  & 96.0  & 85.0  & 70.0 \\
BGE-reranker             & 98.0  & 98.0  & 48.0  & 53.0 \\
\ourmethod (D2DSIM-EBD)  & 69.0  & 22.0  & 10.0  & 10.0 \\
\ourmethod (D2DSIM-BM25) & 34.0  & 15.0  & 2.0   & 2.0  \\
\ourmethod (HRSIM)       & 19.0  & 8.0   & 1.0   & 2.0  \\
\bottomrule
\end{tabular}%
}
\caption{The percentage of poisoned documents in the given context to LLM before and after different defense methods on the MS-MARCO dataset. A lower value is better. Method Keyword not included as it does not conduct reranking.}
\label{tab:MS_rerank}
\end{table*}

\begin{table*}[]
\resizebox{\linewidth}{!}{%
\begin{tabular}{lcccccccc}
\toprule
                          \textbf{Defense Method} & \textbf{PoisonedRAG} & \textbf{PoisonedRAG(Hotflip)} & \textbf{PIA} & \textbf{Phantom} \\
\midrule
No Defense               & 100.0 & 100.0 & 100.0 & 100.0 \\
\midrule
HLATR                    & 100.0 & 100.0 & 100.0 & 99.0 \\
BGE-reranker             & 98.0  & 100.0 & 98.0  & 98.0 \\
\ourmethod (D2DSIM-EBD)  & 84.0  & 66.0  & 52.0  & 49.0 \\
\ourmethod (D2DSIM-BM25) & 64.0  & 53.0  & 35.0  & 32.0 \\
\ourmethod (HRSIM)       & 19.0  & 18.0  & 26.0  & 20.0 \\
\bottomrule
\end{tabular}%
}
\caption{The percentage of poisoned documents in the given context to LLM before and after different defense methods on the HotpotQA dataset. A lower value is better. Method Keyword not included as it does not conduct reranking.}
\label{tab:Hotpot_rerank}
\end{table*}
\clearpage


\begin{table*}[t]
\centering
\resizebox{\linewidth}{!}{%
\begin{tabular}{@{}llccc@{}}
\toprule
\textbf{Model} & \textbf{Defense} & \textbf{HotpotQA} & \textbf{NQ} & \textbf{MS-MARCO} \\
\midrule

\multirow{8}{*}{GPT-4o}        
                                & No-RAG               &  26.0±0.0&  20.0±0.0&  14.0±0.0\\
                                & None               & 60.7±1.2& 58.7±0.5& 65.0±0.0\\
                               & HLATR              &  63.3±2.6&  61.3±0.5&  67.3±0.5\\
                               & BGE-reranker               &  62.7±1.2&  66.0±0.8&  70.3±1.9\\ 
                               & Keyword Aggregation  &  61.7±0.9&  47.0±0.0&  47.0±0.0\\
                               & \ourmethod (D2DSIM-EBD)  &  57.3±0.9&  55.0±0.8&  62.0±0.0\\
                               & \ourmethod (D2DSIM-BM25)    &  56.3±0.9&  58.7±0.5&  66.0±0.0\\
                               & \ourmethod (HRSIM) &  51.0±1.4&  62.0±0.0&  65.0±0.0\\ \midrule

\multirow{8}{*}{Llama3.1-70b-Instruct}  
                                & No-RAG               &  26.0±1.6&  21.3±1.2&  15.3±0.5\\
                                & None               &  60.0±0.8&  66.7±0.5&  53.7±0.5\\
                                & HLATR              &  62.7±0.5&  60.7±0.5&  53.0±0.8\\
                               & BGE-reranker               &  59.0±0.0&  66.0±0.0&  55.7±0.5\\
                               & Keyword Aggregation      &  24.3±1.20&  22.3±0.5&  15.3±0.5\\
                               & \ourmethod (D2DSIM-EBD)  &  47.7±0.5&  55.7±0.5&  46.3±0.5\\
                               & \ourmethod (D2DSIM-BM25)    &  39.0±0.8&  58.3±0.5&  52.7±0.5\\
                               & \ourmethod (HRSIM) &  32.0±0.8&  54.7±0.9&  53.0±0.8\\
                                \midrule
                                
\multirow{8}{*}{Qwen2.5-7b-Instruct}   
                               & No-RAG               &  6.0±0.0&  11.0±0.0&  4.7±0.5\\
                               & None               &  46.0±0.0&  50.7±1.2&  51.0±0.0\\
                               & HLATR              &  47.3±0.5&  48.0±0.0&  44.0±1.4\\
                               & BGE-reranker       &  44.7±0.5&  49.0±0.0&  47.3±0.5\\ 
                               & Keyword Aggregation   &  13.0±0.0&  16.0±0.0&  23.0±0.8\\ 
                               & \ourmethod (D2DSIM-EBD)  &  40.7±0.5&  45.7±0.5&  44.0±0.8\\
                               & \ourmethod (D2DSIM-BM25)    &  37.7±0.5&  48.3±0.5&  46.0±0.8\\
                               & \ourmethod (HRSIM) &  30.0±0.0&  44.7±1.2&  48.7±0.9\\ \midrule

\multirow{8}{*}{Qwen2.5-14b-Instruct}   
                                & No-RAG             & 17.3±0.5 & 17.3±0.5 & 9.3±0.5 \\
                                & None               & 45.7±0.5 & 45.7±0.5 & 45.3±0.5 \\
                                & HLATR              & 34.7±0.5 & 34.7±0.5 & 34.3±0.5 \\
                                & BGE-reranker       & 36.7±1.2 & 36.7±1.2 & 37.3±0.5 \\ 
                                & Keyword Aggregation   & 17.3±0.5 & 17.3±0.5 & 9.7±0.5 \\ 
                                & \ourmethod (D2DSIM-EBD)  & 33.0±0.8 & 33.0±0.8 & 38.7±0.5 \\
                                & \ourmethod (D2DSIM-BM25)    & 43.7±0.5 & 43.7±0.5 & 36.3±0.9 \\
                                & \ourmethod (HRSIM) & 41.7±0.9 & 41.7±0.9 & 39.3±0.5 \\\bottomrule

\end{tabular}
}
\caption{EM scores of defense methods on GPT-4o, Llama3.1-70b-Instruct and Qwen2.5-7b-Instruct when presented
with benign inputs. }
\label{tab:benign_others}
\end{table*}

\begin{table*}[t]
\centering
\resizebox{\linewidth}{!}{%
\begin{tabular}{@{}llccc@{}}
\toprule
\multirow{2}{*}{\textbf{Model}} & \multirow{2}{*}{\textbf{Defense}} & \textbf{HotpotQA} & \textbf{NQ} & \textbf{MS-MARCO} \\
&  & \small{ASR $\downarrow$ / EM $\uparrow$} & \small{ASR $\downarrow$ / EM $\uparrow$} & \small{ASR $\downarrow$ / EM $\uparrow$} \\ \midrule

\multirow{6}{*}{GPT-4o}        & None               & 42.0±0.0 / 40.0±0.8& 29.3±0.9 / 38.0±1.4& 24.0±0.0 / 46.0±0.0\\
                               & HLATR              & 38.0±0.8 / 44.7±2.0& 27.0±0.8 / 48.3±1.2& 21.0±0.0 / 53.0±0.0\\
                               & BGE-reranker               & 37.3±1.2 / 45.7±1.7& 24.7±0.5 / 54.7±0.5& 20.0±0.0 / 54.0±0.0\\ 
                               & Keyword Aggregation  & \colorbox{cellgood}{6.7±0.5 / 58.3±1.9}& \colorbox{cellgood}{1.0±0.0 / 46.0±0.0}& \colorbox{cellgood}{5.7±0.5 / 45.7±0.5}\\ 
                                & \ourmethod (D2DSIM-EBD)  & 37.0±0.0 / 42.0±0.8& 9.7±0.5 / 46.3±0.9& 19.0±0.0 / 51.0±0.0\\
                               & \ourmethod (D2DSIM-BM25)    & 23.7±0.5 / 43.3±0.9& 5.0±0.0 / 60.0±0.0& 10.0±0.0 / 64.0±0.0\\
                               & \ourmethod (HRSIM) & \colorbox{cellgood}{5.0±0.0 / 50.0±1.4}& \colorbox{cellgood}{1.0±0.0 / 64.0±1.4}& \colorbox{cellgood}{4.0±0.0 / 67.0±0.0}\\\midrule

\multirow{6}{*}{Llama3.1-70b-Instruct}  & None               & 57.7±0.9 / 37.3±0.9& 56.7±0.5 / 29.7±0.5& 54.3±1.2 / 28.3±0.9\\
                               & HLATR              & 53.0±0.8 / 43.3±0.5& 49.0±0.0 / 39.0±0.0& 40.7±1.2 / 37.0±0.8\\
                               & BGE-reranker              & 53.3±0.5 / 41.7±0.5& 49.3±0.5 / 38.0±0.0& 37.0±0.8 / 37.0±0.8\\
                               & Keyword Aggregation  & \colorbox{cellgood}{4.7±0.5 / 26.0±0.8}& \colorbox{cellgood}{3.0±0.0 / 22.3±0.5}& \colorbox{cellgood}{3.0±0.0 / 58.0±2.2}\\
                               & \ourmethod (D2DSIM-EBD)  & 45.3±0.5 / 37.3±0.5& 26.0±0.0 / 44.0±0.0& 34.3±0.5 / 39.3±0.5\\
                               & \ourmethod (D2DSIM-BM25)    & 36.0±0.0 / 37.7±0.5& 11.0±0.0 / 56.3±0.5& 15.0±0.0 / 50.7±0.5\\
                               & \ourmethod (HRSIM) & \colorbox{cellgood}{8.3±0.5 / 37.7±0.5}& \colorbox{cellgood}{2.7±0.5 / 53.0±0.0}& \colorbox{cellgood}{9.0±0.0 / 54.0±0.0}\\ \midrule
 
\multirow{6}{*}{Qwen2.5-7b-Instruct}    & None              & 62.0±0.0 / 24.0±0.0& 50.3±0.5 / 26.7±0.5& 49.0±0.0 / 28.7±0.5\\
                               & HLATR              & 60.0±0.0 / 28.7±0.5& 42.7±0.5 / 31.3±0.5& 41.0±0.0 / 28.0±0.8\\
                               & BGE-reranker       & 60.0±0.0 / 30.7±0.5& 47.7±0.5 / 29.3±0.5& 42.0±0.8 / 29.3±0.5\\ 
                               & Keyword Aggregation & \colorbox{cellgood}{4.7±0.5 / 6.0±0.0}& \colorbox{cellgood}{3.0±0.0 / 11.0±0.0}& \colorbox{cellgood}{9.3±0.9 / 25.0±0.8}\\ 
                               & \ourmethod (D2DSIM-EBD)  & 57.0±0.0 / 24.3±0.5& 24.3±0.5 / 35.3±1.2& 37.3±0.5 / 31.3±0.5\\
                               & \ourmethod (D2DSIM-BM25)   & 42.3±0.5 / 27.7±0.5& 12.7±0.5 / 45.0±0.8& 23.7±0.5 / 38.0±1.4\\
                               & \ourmethod (HRSIM)& \colorbox{cellgood}{7.7±0.5 / 34.0±0.8}& \colorbox{cellgood}{5.3±0.5 / 41.0±0.0}& \colorbox{cellgood}{12.3±0.5 / 39.0±0.8}\\ \midrule

\multirow{7}{*}{Qwen2.5-14b-Instruct}   
                                & None               & 47.7±0.5 / 17.0±0.0 & 43.3±0.5 / 12.3±0.5 & 38.0±0.0 / 19.7±0.5 \\
                                & HLATR              & 42.7±0.5 / 17.7±0.5 & 36.3±0.5 / 19.7±1.2 & 26.7±0.5 / 23.3±1.2 \\
                                & BGE-reranker       & 43.0±0.0 / 19.7±0.5 & 35.3±0.5 / 23.7±0.5 & 25.0±0.0 / 22.7±0.5 \\ 
                                & Keyword Aggregation   & \colorbox{cellgood}{4.0±0.0 / 22.3±0.5} & \colorbox{cellgood}{5.0±0.0 / 18.3±0.5} & \colorbox{cellgood}{3.0±0.0 / 9.7±0.5} \\ 
                                & \ourmethod (D2DSIM-EBD)  & 34.0±0.0 / 16.0±0.0 & 13.0±0.0 / 27.7±1.2 & 26.0±0.0 / 27.0±0.0 \\
                                & \ourmethod (D2DSIM-BM25)    & 28.0±0.0 / 20.0±1.4 & \colorbox{cellgood}{5.0±0.0 / 38.3±0.5} & 13.3±0.5 / 33.7±0.9 \\
                                & \ourmethod (HRSIM) & \colorbox{cellgood}{7.3±0.5 / 21.3±0.9} & \colorbox{cellgood}{1.0±0.0 / 40.0±0.0} & \colorbox{cellgood}{8.3±0.5 / 36.3±1.2} \\\bottomrule

\end{tabular}
}
\caption{ASR and EM (\%) for various defense methods on PoisonedRAG on GPT-4o, Llama3.1-70b-Instruct and Qwen2.5-7b-Instruct. \colorbox{cellgood}{\textbf{Blue}} cells indicate top-two lowest ASR.}
\label{tab:other_poisoned}
\end{table*}

\begin{table*}[t]
\centering
\resizebox{\linewidth}{!}{%
\begin{tabular}{@{}llccc@{}}
\toprule
\multirow{2}{*}{\textbf{Model}} & \multirow{2}{*}{\textbf{Defense}} & \textbf{HotpotQA} & \textbf{NQ} & \textbf{MS-MARCO} \\
&  & \small{ASR $\downarrow$ / EM $\uparrow$} & \small{ASR $\downarrow$ / EM $\uparrow$} & \small{ASR $\downarrow$ / EM $\uparrow$} \\ \midrule

\multirow{6}{*}{GPT-4o}        & None               & 45.3±0.5 / 41.7±0.5& 32.3±0.5 / 39.0±1.4& 24.7±0.9 / 46.0±1.4\\
                               & HLATR              & 42.0±0.8 / 45.0±0.8& 28.3±0.9 / 48.7±1.9& 19.7±2.4 / 53.0±1.4\\
                               & BGE-reranker              & 40.0±0.0 / 41.3±0.5& 27.0±0.0 / 49.0±0.0& 20.0±0.0 / 53.7±0.9\\ 
                               & Keyword Aggregation  & \colorbox{cellgood}{8.7±0.5 / 59.3±1.9}& \colorbox{cellgood}{1.0±0.0 / 46.0±0.0}& \colorbox{cellgood}{4.0±0.0 / 48.0±1.4}\\ 
                                & \ourmethod (D2DSIM-EBD)  & 31.7±0.5 / 45.3±1.2& 5.0±0.8 / 55.3±1.2& 11.3±0.5 / 56.0±1.4\\
                               & \ourmethod (D2DSIM-BM25)    & 21.0±0.0 / 46.3±0.9& 5.0±0.0 / 61.3±0.5& 7.3±0.5 / 67.0±1.4\\
                               & \ourmethod (HRSIM) & \colorbox{cellgood}{5.0±0.0 / 49.3±1.9}& \colorbox{cellgood}{1.0±0.0 / 63.3±2.4}& \colorbox{cellgood}{4.0±0.0 / 66.3±0.5}\\\midrule

\multirow{6}{*}{Llama3.1-70b-Instruct}  & None               & 56.7±0.9 / 33.3±0.9& 54.7±2.1 / 26.7±1.7& 47.7±0.5 / 29.0±0.8\\
                               & HLATR              & 52.0±2.2 / 37.3±1.2& 47.3±2.1 / 35.7±0.9& 32.3±0.5 / 37.0±0.8\\
                               & BGE-reranker              & 48.3±1.2 / 44.3±1.9& 42.7±1.2 / 41.3±1.2& 35.7±1.9 / 33.3±0.9\\
                               & Keyword ~\cite{xiang2024certifiablyrobustragretrieval}         & \colorbox{cellgood}{4.7±0.5 / 26.0±0.8}& \colorbox{cellgood}{3.0±0.0 / 22.0±0.0}& \colorbox{cellgood}{3.0±0.0 / 57.0±0.8}\\
                               & \ourmethod (D2DSIM-EBD)  & 37.0±0.0 / 40.3±1.7& 11.0±1.6 / 48.7±2.1& 15.3±1.7 / 45.7±0.5\\
                               & \ourmethod (D2DSIM-BM25)    & 33.3±0.9 / 37.7±2.1& 6.7±0.5 / 56.0±0.8& 10.3±0.5 / 51.7±1.2\\
                               & \ourmethod (HRSIM) & \colorbox{cellgood}{8.7±0.5 / 36.7±2.6}& \colorbox{cellgood}{1.0±0.0 / 54.0±0.8}& \colorbox{cellgood}{6.7±0.5 / 52.3±2.4}\\ \midrule

\multirow{6}{*}{Qwen2.5-7b-Instruct}    & None              & 58.7±0.9 / 30.7±1.2& 58.0±2.2 / 22.3±2.1& 51.0±1.4 / 31.3±2.9\\
                               & HLATR              & 55.7±0.9 / 33.7±2.1& 51.0±0.0 / 29.0±0.8& 36.3±3.3 / 33.0±3.3\\
                               & BGE-reranker       & 54.0±1.6 / 33.7±2.6& 51.0±0.8 / 29.3±0.5& 37.3±4.0 / 33.3±3.9\\ 
                               & Keyword     & \colorbox{cellgood}{4.7±0.5 / 6.0±0.0}& \colorbox{cellgood}{3.0±0.0 / 11.0±0.0}& \colorbox{cellgood}{10.3±0.5 / 23.7±0.5}\\ 
                               & \ourmethod (D2DSIM-EBD)  & 45.7±0.9 / 31.0±1.6& 14.7±1.7 / 41.0±3.6& 19.0±1.6 / 36.3±0.5\\
                               & \ourmethod (D2DSIM-BM25)   & 38.3±0.5 / 31.7±1.2& 12.0±2.2 / 42.0±0.8& 14.7±1.2 / 40.7±0.5\\
                               & \ourmethod (HRSIM) & \colorbox{cellgood}{6.0±0.0 / 33.0±0.0}& \colorbox{cellgood}{4.3±0.9 / 45.3±0.5}& \colorbox{cellgood}{10.7±0.5 / 39.0±1.4}\\ \midrule

\multirow{7}{*}{Qwen2.5-14b-Instruct}   
                                & None               & 50.0±1.6 / 20.0±0.8 & 47.3±2.4 / 16.0±0.8 &42.3±0.9 / 20.0±2.2   \\
                                & HLATR              & 47.3±1.2 / 17.3±1.2 & 37.0±0.8 / 21.0±2.2 &28.3±1.2 / 28.0±1.6   \\
                                & BGE-reranker       & 43.3±2.9 / 23.7±1.7 & 32.7±0.5 / 27.7±2.9 &27.3±0.5 / 25.7±1.2   \\ 
                                & Keyword Aggregation   & \colorbox{cellgood}{4.3±0.5 / 22.3±0.9} & 5.0±0.0 / 16.7±0.5 &\colorbox{cellgood}{3.0±0.0 / 10.0±0.0}   \\ 
                                & \ourmethod (D2DSIM-EBD)  & 34.0±2.9 / 17.3±0.5  & 10.3±0.5 / 28.0±1.6 &15.0±0.8 / 34.3±2.1   \\
                                & \ourmethod (D2DSIM-BM25)    & 31.3±1.2 / 22.7±0.9 & \colorbox{cellgood}{4.3±1.2 / 39.7±1.7} & 12.0±0.8 / 36.0±2.2   \\
                                & \ourmethod (HRSIM) & \colorbox{cellgood}{9.0±0.0 / 22.0±0.0} & \colorbox{cellgood}{1.3±0.5 / 42.3±0.5} &\colorbox{cellgood}{7.7±0.9 / 37.0±0.8}  \\\bottomrule

\end{tabular}
}
\caption{ASR and EM (\%) for various defense methods on PoisonedRAG(Hotflip). \colorbox{cellgood}{\textbf{Blue}} cells indicate top-two lowest ASR.}
\label{tab:other_hotflip}
\end{table*}

\begin{table*}[t]
\centering
\resizebox{\linewidth}{!}{%
\begin{tabular}{@{}llccc@{}}
\toprule
\multirow{2}{*}{\textbf{Model}} & \multirow{2}{*}{\textbf{Defense}} & \textbf{HotpotQA} & \textbf{NQ} & \textbf{MS-MARCO} \\
&  & \small{ASR $\downarrow$ / EM $\uparrow$} & \small{ASR $\downarrow$ / EM $\uparrow$} & \small{ASR $\downarrow$ / EM $\uparrow$} \\ \midrule

\multirow{6}{*}{GPT-4o}        & None               & 99.0±0.0 / 0.3±0.5& 95.7±0.5 / 3.7±0.5& 80.0±0.0 / 11.0±0.0\\
                               & HLATR              & 97.6±0.9 / 1.3±0.9& 78.0±0.0 / 15.0±0.0& 53.0±0.0 / 32.0±0.0\\
                               & BGE-reranker       & 87.3±0.5 / 7.0±1.4& 36.0±1.4 / 39.7±0.9& 24.0±0.0 / 51.0±0.0\\ 
                               & Keyword Aggregation  & \colorbox{cellgood}{0.0±0.0 / 53.7±2.4}& \colorbox{cellgood}{0.0±0.0 / 44.0±0.0}& \colorbox{cellgood}{0.0±0.0 / 45.7±0.5}\\ 
                                & \ourmethod (D2DSIM-EBD)  & 30.7±0.5 / 42.3±0.9& 2.0±0.0 / 57.3±0.5& 2.0±0.0 / 60.0±0.0\\
                               & \ourmethod (D2DSIM-BM25)   & 40.0±1.4 / 36.3±0.9& 10.7±0.9 / 57.3±0.9& 0.0±0.0 / 68.0±0.0\\
                               & \ourmethod (HRSIM) & \colorbox{cellgood}{25.0±0.0 / 42.7±0.5}& \colorbox{cellgood}{1.0±0.0 / 63.7±0.9}& \colorbox{cellgood}{0.0±0.0 / 68.0±0.0}\\\midrule

\multirow{6}{*}{Llama3.1-70b-Instruct}  & None               & 100.0±0.0 / 0.0±0.0& 98.0±0.0 / 2.0±0.0& 88.0±0.0 / 8.0±0.0\\
                               & HLATR              & 100.0±0.0 / 0.0±0.0& 91.7±0.5 / 5.3±0.5& 84.0±0.0 / 8.7±0.5\\
                               & BGE-reranker       & 98.0±0.0 / 2.0±0.0& 42.3±0.5 / 38.7±0.5& 43.0±0.0 / 30.3±0.5\\
                               & Keyword Aggregation  & \colorbox{cellgood}{0.0±0.0 / 26.7±0.5}& \colorbox{cellgood}{0.0±0.0 / 23.0±1.4}& \colorbox{cellgood}{0.0±0.0 / 59.3±0.9}\\
                               & \ourmethod (D2DSIM-EBD)  & 33.0±0.0 / 29.0±0.0& 2.0±0.0 / 55.3±0.5& 3.0±0.0 / 49.0±0.8\\
                               & \ourmethod (D2DSIM-BM25)    & 42.0±0.0 / 25.0±0.0& 12.0±0.0 / 52.0±0.8& 2.0±0.0 / 54.3±1.2\\
                               & \ourmethod (HRSIM) & \colorbox{cellgood}{26.0±0.0 / 32.0±0.8}& \colorbox{cellgood}{1.3±0.5 / 55.3±0.5}& \colorbox{cellgood}{1.0±0.0 / 54.7±0.5}\\ \midrule

\multirow{6}{*}{Qwen2.5-7b-Instruct}    & None              & 5.3±0.5 / 22.7±0.5& 5.7±0.5 / 17.0±0.0& 6.0±0.0 / 27.0±0.8\\
                               & HLATR             & 14.0±0.8 / 24.0±1.4& 17.7±0.9 / 12.7±0.9& 18.0±0.0 / 20.7±0.5\\
                               & BGE-reranker      & 25.0±0.0 / 17.0±0.0& 23.0±0.0 / 31.7±0.5& 18.3±0.5 / 32.0±0.0\\ 
                               & Keyword Aggregation    & \colorbox{cellgood}{0.0±0.0 / 6.0±0.0}& \colorbox{cellgood}{0.0±0.0 / 11.0±0.0}& \colorbox{cellgood}{0.0±0.0 / 21.3±0.5}\\ 
                               & \ourmethod (D2DSIM-EBD)  & 12.0±0.0 / 34.7±0.9& 2.0±0.0 / 47.0±0.8& 3.0±0.0 / 41.7±0.5\\
                               & \ourmethod (D2DSIM-BM25)   & 15.0±0.0 / 28.0±0.0& 8.0±0.0 / 43.7±0.5& \colorbox{cellgood}{1.0±0.0 / 44.0±1.4}\\
                               & \ourmethod (HRSIM) & \colorbox{cellgood}{8.7±0.5 / 35.3±0.5}& \colorbox{cellgood}{2.0±0.0 / 46.3±0.9}& \colorbox{cellgood}{1.0±0.0 / 47.3±1.7}\\ \midrule

\multirow{7}{*}{Qwen2.5-14b-Instruct}   
                                & None               & 99.0±0.0 / 0.0±0.0 & 94.0±0.0 / 3.0±0.0 & 87.0±0.0 / 6.7±0.5 \\
                                & HLATR              & 98.0±0.0 / 0.0±0.0 & 88.7±0.5 / 3.0±0.0 & 83.0±0.0 / 5.7±0.5 \\
                                & BGE-reranker       & 98.0±0.0 / 1.3±0.5 & 42.0±0.0 / 21.3±0.5 & 43.0±0.0 / 21.3±0.5 \\ 
                                & Keyword Aggregation   & \colorbox{cellgood}{0.0±0.0 / 23.0±0.0} & \colorbox{cellgood}{0.0±0.0 / 18.7±0.5} & \colorbox{cellgood}{0.0±0.0 / 9.7±0.5} \\ 
                                & \ourmethod (D2DSIM-EBD)  & 33.0±0.0 / 14.0±0.0 & 2.0±0.0 / 29.0±1.4 & 3.0±0.0 / 37.3±0.5 \\
                                & \ourmethod (D2DSIM-BM25)    & 40.7±0.5 / 17.0±0.0 & 12.0±0.0 / 35.3±0.5 & 2.0±0.0 / 37.3±0.5 \\
                                & \ourmethod (HRSIM) &\colorbox{cellgood}{27.0±0.0 / 18.0±0.0} & \colorbox{cellgood}{2.0±0.0 / 37.7±0.9} & \colorbox{cellgood}{1.0±0.0 / 40.3±0.5} \\\bottomrule

\end{tabular}
}
\caption{ASR and EM (\%) for various defense methods on PIA. \colorbox{cellgood}{\textbf{Blue}} cells indicate top-two lowest ASR.}
\label{tab:other_PIA}
\end{table*}

\begin{table*}[t]
\centering
\resizebox{\linewidth}{!}{%
\begin{tabular}{@{}llccc@{}}
\toprule
\multirow{2}{*}{\textbf{Model}} & \multirow{2}{*}{\textbf{Defense}} & \textbf{HotpotQA} & \textbf{NQ} & \textbf{MS-MARCO} \\
&  & \small{ASR $\downarrow$ / EM $\uparrow$} & \small{ASR $\downarrow$ / EM $\uparrow$} & \small{ASR $\downarrow$ / EM $\uparrow$} \\ \midrule

\multirow{6}{*}{GPT-4o}        & None               & 57.3±0.5 / 25.3±0.9& 37.0±0.0 / 18.7±0.5& 21.0±0.0 / 45.0±0.0\\
                               & HLATR              & 47.0±1.4 / 27.3±0.5& 36.3±0.5 / 22.3±0.5& 18.0±0.0 / 53.0±0.0\\
                               & BGE-reranker       & 35.7±0.9 / 29.7±0.5& 20.3±0.5 / 32.3±0.5& 19.0±0.0 / 53.0±0.0\\ 
                               & Keyword Aggregation  & \colorbox{cellgood}{0.0±0.0 / 57.0±0.0}& \colorbox{cellgood}{0.0±0.0 / 48.0±0.0}& \colorbox{cellgood}{0.0±0.0 / 45.0±0.0}\\ 
                                & \ourmethod (D2DSIM-EBD)  & 30.0±1.4 / 35.3±0.5& 3.7±0.5 / 43.3±1.9& 2.0±0.0 / 53.0±0.0\\
                               & \ourmethod (D2DSIM-BM25)   & 7.3±0.9 / 40.0±1.4& 2.0±0.0 / 51.0±0.0& 0.3±0.5 / 63.0±0.0\\
                               & \ourmethod (HRSIM) & \colorbox{cellgood}{3.3±0.9 / 41.3±0.9}& \colorbox{cellgood}{0.0±0.0 / 50.0±1.4}& \colorbox{cellgood}{0.0±0.0 / 63.7±0.5}\\\midrule

\multirow{6}{*}{Llama3.1-70b-Instruct}  & None               & 98.7±0.5 / 1.3±0.5& 90.7±1.2 / 7.3±1.2& 74.3±1.2 / 19.7±0.5\\
                               & HLATR              & 98.0±0.8 / 0.7±0.5& 93.7±0.9 / 5.3±0.5& 78.0±1.6 / 13.3±0.9\\
                               & BGE-reranker       & 96.3±0.5 / 3.7±0.5& 75.7±0.9 / 14.0±0.8& 70.7±0.9 / 20.3±1.7\\
                               & Keyword Aggregation & \colorbox{cellgood}{0.0±0.0 / 18.7±0.5}& \colorbox{cellgood}{0.0±0.0 / 17.3±0.5}& \colorbox{cellgood}{0.0±0.0 / 51.3±1.2}\\
                               & \ourmethod (D2DSIM-EBD)  & 60.3±2.9 / 16.3±1.2& 12.7±2.6 / 41.7±1.7& 13.7±2.4 / 45.3±2.1\\
                               & \ourmethod (D2DSIM-BM25)    & 27.0±1.4 / 27.7±1.2& 5.3±0.5 / 49.3±0.5& 1.3±0.5 / 55.3±0.5\\
                               & \ourmethod (HRSIM) & \colorbox{cellgood}{11.3±0.9 / 27.3±1.2}& \colorbox{cellgood}{0.7±0.5 / 50.7±1.2}& \colorbox{cellgood}{0.0±0.0 / 56.0±0.8}\\ \midrule

\multirow{6}{*}{Qwen2.5-7b-Instruct}    & None              & 58.7±3.8 / 18.3±1.2& 56.0±2.9 / 12.0±2.2& 40.0±1.4 / 25.3±2.1\\
                               & HLATR             & 63.0±1.4 / 17.7±2.1& 71.0±1.6 / 9.3±1.7& 48.3±2.1 / 18.7±2.1\\
                               & BGE-reranker      & 62.3±4.1 / 19.7±0.5& 57.7±2.6 / 19.7±1.2& 50.3±0.9 / 25.7±2.5\\ 
                               & Keyword Aggregation    & \colorbox{cellgood}{0.0±0.0 / 1.0±0.0}& \colorbox{cellgood}{0.0±0.0 / 5.0±0.0}& \colorbox{cellgood}{0.0±0.0 / 5.0±0.0}\\ 
                               & \ourmethod (D2DSIM-EBD)  & 41.0±2.8 / 17.0±3.7& 11.0±2.8 / 32.0±0.8& 11.7±1.7 / 40.7±2.1\\
                               & \ourmethod (D2DSIM-BM25)  & 24.0±0.0 / 27.7±2.1& 5.3±0.5 / 35.3±1.2& 0.3±0.5 / 45.7±0.9\\
                               & \ourmethod (HRSIM) & \colorbox{cellgood}{14.0±2.4 / 27.3±0.9}& \colorbox{cellgood}{0.7±0.5 / 36.3±0.5}& \colorbox{cellgood}{0.0±0.0 / 48.7±1.7}\\ \midrule

\multirow{7}{*}{Qwen2.5-14b-Instruct}   
                                & None               & 67.7±2.1 / 0.3±0.5 & 51.0±1.4 / 4.0±1.6 & 43.7±2.5 / 16.3±2.1 \\
                                & HLATR              & 72.0±2.9 / 1.0±0.0 & 56.0±2.2 / 3.7±0.9 & 52.3±3.4 / 12.7±0.5 \\
                                & BGE-reranker       & 81.7±2.9 / 1.0±0.8 & 58.7±1.2 / 11.3±1.2 & 51.0±2.2 / 17.0±0.8 \\ 
                                & Keyword Aggregation   & \colorbox{cellgood}{0.0±0.0 / 12.7±0.5} & \colorbox{cellgood}{0.0±0.0 / 11.3±1.2} &  \colorbox{cellgood}{0.0±0.0 / 7.0±0.0} \\ 
                                & \ourmethod (D2DSIM-EBD)  & 44.7±3.1 / 5.0±0.8 & 12.0±2.2 / 24.7±2.5 & 11.0±2.2 / 34.7±0.5 \\
                                & \ourmethod (D2DSIM-BM25)    & 23.7±0.9 / 18.7±0.5 & 5.0±0.8 / 34.0±1.4 & 0.3±0.5 / 38.0±0.8 \\
                                & \ourmethod (HRSIM) & \colorbox{cellgood}{7.7±1.2 / 16.3±0.5} & \colorbox{cellgood}{0.0±0.0 / 34.7±0.9} & \colorbox{cellgood}{0.0±0.0 / 42.3±0.5} \\\bottomrule

\end{tabular}
}
\caption{ASR and EM (\%) for various defense methods on Phantom. \colorbox{cellgood}{\textbf{Blue}} cells indicate top-two lowest ASR.}
\label{tab:other_phantom}
\end{table*}

\end{document}